\documentclass[12pt,a4paper]{article}

\usepackage{a4}
\usepackage{amsfonts}
\usepackage{amssymb}
\usepackage{epsfig}
\usepackage{psfig}
\usepackage{psfrag}
\usepackage{dsfont}

\newcommand{\beqn}{\begin{eqnarray}}
\newcommand{\eeqn}{\end{eqnarray}}
\newcommand{\be}{\begin{equation}}
\newcommand{\ee}{\end{equation}}
\newcommand{\non}{\nonumber \\}

\newcommand{\tg}{{\tilde G}}
\newcommand{\tga}{{\tilde \gamma}}

\begin{document}

\title{}
\begin{flushright}
\vspace{-3cm}
{\small SPIN-03/08 \\
        ITP-UU-03/14 \\
        MIT-CTP-3366 \\
        ROM2F/2003/06 \\
        hep-th/0305183}
\end{flushright}
\vspace{1cm}

\begin{center}
{\Large\bf An Orientifold with Fluxes and Branes \\[.3cm] via T-duality}
\end{center}

\vspace{0.5cm}

\author{}
\date{}
\thispagestyle{empty}

\begin{center}

{\bf Marcus Berg,}\footnote{e-mail: berg@roma2.infn.it}$^{,\dag}$
{\bf Michael Haack}\footnote{e-mail: haack@roma2.infn.it}$^{,\dag}$
{\bf and Boris K\"ors}\footnote{e-mail: kors@lns.mit.edu}$^{,*}$
\vspace{0.5cm}

{\it
$^\dag$ Dipartimento di Fisica \\
Universit\`a di Roma, Tor Vergata \\
00133 Rome, Italy
}


\hbox{
\parbox{7cm}{
\begin{center}
{\it
$^*$Spinoza Institute and \\
Institute for Theoretical Physics \\
Utrecht University \\
{{Utrecht, The Netherlands}} \\
}
\end{center}
}
\hspace{-.5cm}
\parbox{8cm}{
\begin{center}
{\it

$^*$Center for Theoretical Physics \\
Laboratory for Nuclear Science \\
and Department of Physics \\
Massachusetts Institute of Technology \\
Cambridge, Massachusetts 02139, USA \\
}
\end{center}
}
}

\end{center}

\begin{center}
{\bf Abstract} \\
\end{center}
String compactifications with non-abelian gauge fields localized
on D-branes, with background NSNS and RR 3-form fluxes, and with
non-trivial warp factors, can naturally exist within T-dual
versions of type I string theory. We develop a systematic
procedure to construct the effective bosonic Lagrangian of type I
T-dualized along a six-torus, including the coupling
to gauge multiplets on D3-branes and the modifications due to
3-form fluxes. Looking for solutions to the ten-dimensional equations of motion,
we find warped products of Minkowski space and Ricci-flat internal
manifolds. Once the warp factor is neglected, the resulting no-scale 
scalar potential of the effective four-dimensional theory combines 
those known for 3-form fluxes and for internal Yang-Mills fields and 
stabilizes many of the moduli. We perform an explicit comparison of our 
expressions to those obtained from ${\cal N}=4$ gauged supergravity 
and find agreement. We also comment on
the possibility to include D9-branes with world-volume gauge
fluxes in the background with 3-form fluxes.

\clearpage

\section{Introduction}
\setcounter{footnote}{0}

Orientifolds  provide
a natural framework for string compactifications that can accommodate space-time
filling D-branes, internal fluxes for the various tensor field strengths and
non-trivial warp factors at the same time.\footnote{See \cite{hepth0204089} for a comprehensive
introduction to orientifolds.} Since these are the main ingredients
in many of the recent phenomenological investigations concerning low string scale
models \cite{Arkani-Hamed:1998rs,Antoniadis:1998ig},
moduli stabilization through flux-induced potentials \cite{Polchinski:1995sm}-
\cite{Becker:2003gq},
Randall-Sundrum-like warped compactifications
\cite{Randall:1999ee,Randall:1999vf} or any kind of brane world
scenarios, they are of central interest among the classes of
string compactifications relevant for four-dimensional particle
physics. They evade the powerful no-go theorems prohibiting warped
compactifications with fluxes to four-dimensional Minkowski space
\cite{deWit:1986xg,Maldacena:2000mw,Ivanov:2000fg,Gauntlett:2002sc}
within the context of Kaluza-Klein
reduction of supergravity, since in contrast to traditional supergravity theories
they contain objects of negative energy density like orientifold planes. However, type I
models, or their T-dual descriptions usually called type I$'$, are notoriously
difficult to treat: no explicit description of the effective action of type I$'$ models has
been given.\footnote{But for some information on ${\cal N}=1$ effective four-dimensional type I supergravity
in a T-duality invariant formulation see
\cite{Ibanez:1998rf}.} The main purpose of the present paper is to remedy this latter
point and provide the effective bosonic action for a certain theory T-dual to
type I string theory, including the coupling of the supergravity fields to the
non-abelian gauge theory sector localized on D3-branes,
as well as the relevant modifications due to background 3-form fluxes. More
precisely, we construct the type I$'$ model with 3-form fluxes, which in the absence of 3-form fluxes
is dual to type I via six T-dualities. We consider this the simplest version of an orientifold
with 3-form fluxes and D-branes and a starting point for phenomenologically more
sophisticated constructions. \\

The motivation to combine models with D-branes and orientifold planes (O-planes) with background 3-form
fluxes comes from the fact that the
former provide interesting non-abelian gauge fields, potentially
with chiral charged matter,\footnote{For the most part of this paper we only consider
D3-branes at smooth points of the background, which do not lead to chiral spectra. Generalizations
of the present model would need to include higher-dimensional branes with non-trivial gauge
fluxes, singular geometries or with point-like intersections to support chiral fermionic zero modes
(see e.g. \cite{Berkooz:1996km,Angelantonj:1996uy,Aldazabal:1998mr}), a
possibility which we also comment on in section \ref{sec6}.}
while the latter add to the scalar
potential of the effective theory, such that at least some scalar fields
get massive and decouple.
This removes some of the vacuum degeneracy, a necessary step on the way to
realistic string models.
The total scalar potential gets a contribution not only from the
ten-dimensional kinetic term of the type IIB 3-form, projected to type I$'$ and
including a Chern-Simons (CS) correction, but also from the Dirac-Born-Infeld (DBI)
action of the Yang-Mills (YM) gauge fields supported by the D-branes.
Both the 3-form flux and the gauge fields
add positive energy densities and the constraints that follow from the equations
of motion require this energy density to cancel against the negative contribution
of orientifold planes in a Minkowski vacuum. \\

Within supergravity, scalar potentials can be generated by gauging isometries of
the scalar manifold (for a recent review see \cite{deWit:2002vz}).
A systematic procedure can be applied to derive the form of the effective action.
The effective actions obtained from
string compactifications with fluxes have so far been understood as corresponding
to gaugings of a particular type, namely abelian gaugings of Peccei-Quinn (PQ)
isometries, i.e.\ gaugings of axionic shifts. The ${\cal N}=4$ gauged supergravity 
describing the type I$'$ model at hand was studied in a series of papers \cite{TsZi,ADFL,hepth0206241,
hepth0207135,hepth0211027,Andrianopoli:2002vy,D'Auria:2003jk}.
Here we derive the bosonic part of the effective action by explicitly performing
the six T-dualities of type I and make contact with the formulas of \cite{D'Auria:2003jk},
that contains the most explicit formulation of the model, thus
elucidating the correspondence between the field variables used in the
supergravity literature and those naturally arising in string theory.
A deeper understanding of the relation between string
compactifications in non-trivial backgrounds and the gaugings
of the relevant supergravity is certainly worthwhile. \\

It is important to note that for our present purposes T-duality
always manifests itself as a set of field redefinitions in the
action. We are not going to transform a given vacuum of type I,
but instead we transform the action for the dynamical degrees of
freedom of type I string theory into a formulation in terms of the
type I$'$ fields after six T-dualities. In performing the six
T-dualities we have to assume the internal space has six
isometries (i.e.\ it is a six-torus). However, after performing
the transformations one can reinterpret the resulting
four-dimensional action as coming from a non-covariant
ten-dimensional action of the type I$'$ theory. In the
ten-dimensional Lagrangian, the closed string fields are allowed
to depend also on all the internal coordinates, while the open
string fields only
depend on world-volume coordinates (compare the discussion in \cite{hepth9910053}). \\

In deriving the full ten-dimensional action and the correct
equations of motion, we have to make use of an important
difference between type I theory and its T-dual version type I$'$.
This difference permits us, for the type I$'$ theory, to include
complex 3-form fluxes and other non-dynamical backgrounds even
though their fluctuations are projected out. This comes about by
noting that the world sheet parity $\Omega$ which is ``divided
out'' in getting from type IIB to type I string theory is mapped
to $\Omega\Theta(-1)^{F_L}$ under T-duality. Here $\Theta$ is the
reflection of all internal directions $x^i$ and $F_L$ the
left-moving space-time fermion number operator. A field of type
IIB that was odd under $\Omega$ has to vanish at any point in type
I theory, whereas a field odd (even) under
$\Omega\Theta(-1)^{F_L}$ is only required to be antisymmetric
(symmetric) under $x^i \mapsto -x^i$ in type I$'$ theory. The
zero-mode fluctuations of the bosonic fields are constant on a
torus and therefore have to be even under
$\Omega\Theta(-1)^{F_L}$. On the other hand,
one can keep all kinds of background terms, which naively do not
appear in the action for the fluctuations obtained from the type I
action by T-duality. Thus, even though the fluctuations of the
internal components of the two antisymmetric 2-tensors of IIB are
projected out, one can include a background flux for the two
3-form field strengths (which are even under
$\Omega\Theta(-1)^{F_L}$, such that their corresponding potentials
are odd), generating a potential for the closed string moduli.
This is the reason why one is interested in the T-dual version of
type I in the first place. In addition also the odd component of
the field strength $F_5$ with one or five internal indices can
take on a background value as long as it is antisymmetric under
$x^i \mapsto -x^i$. This is important to find explicit solutions
to the ten-dimensional equations of motion as a non-trivial
profile for $C_4$ is needed for
D3-branes or O3-planes to exist. \\

In order to derive the action of type I$'$ including the non-dynamical background terms
in the sense explained above,
we combine two different strategies. First we employ six T-dualities of the type I action.
This produces an effective action equivalent to the
type I action and therefore including all the consistent couplings
of bulk supergravity and non-abelian gauge fields.
However, as we already mentioned, this requires the fields to be independent of the internal coordinates.
In particular we cannot include any torsion for the metric and therefore it is impossible
to directly derive the effects of NSNS 3-form flux in the T-dual theory this way.\footnote{T-duality
in the presence of NSNS flux and the corresponding non-trivial metric configurations have been
discussed in \cite{hepth9912236,DaHu,GLMW,KSTT} (see also
\cite{Strominger:uh,DRS,BD,Cardoso:2002hd,Becker:2003yv,Becker:2003gq} for analogous works on the non-K\"ahler vacua of the
heterotic string). However, there T-duality is not performed along all the internal coordinates.}
We then add in modifications due to fluxes for the NSNS 3-form and $F_5$.
The fact that this adding in by hand is not arbitrary and can be performed in a well defined and
systematic manner rests on the twofold definition of type I$'$:
\be \label{modwsp}
{{\rm T-duality\ of} \atop {\rm Type\ I}} \quad\longleftrightarrow\quad
\frac{\rm Type\ IIB}{\Omega \Theta (-1)^{F_L}}\ .
\ee
This means the closed string sector of the theory can be deduced from the fact that
it is given by projecting the effective action of
type IIB by the T-dual world sheet parity.
This second definition does not give any prescription how to incorporate the
gauge fields, but combined with the knowledge of the T-dual action of type I, it provides
enough additional information to include the closed string
background fields. Thus, the T-duality of type I and the T-dual projection of type IIB are complementary
regarding terms involving the open string fields and the non-dynamical background fields such as the
NSNS 3-form flux, respectively.\footnote{In \cite{Uranga:2002vk} the effect of fluxes in 
the world volume theory of branes has been analyzed from a different point of view, resting on anomaly 
constraints.}
The resulting action then serves as a four-dimensional effective action for a comparison to
the Lagrangian of the gauged ${\cal N}=4$ supergravity as mentioned already. Reinterpreted as
ten-dimensional, we can use it to study more general vacua than torus compactifications as well. \\

The solutions for the ten-dimensional equations of motion
generalize the situation without gauge fields considered e.g.\ in
\cite{deWit:1986xg,hepth0105097}. As in that case, one can find an
explicit solution in the form of a warped product of
four-dimensional Minkowski space and a Ricci-flat internal metric
involving also a non-trivial profile for the components of $F_5$
that respect four-dimensional Poincar\'e invariance. The
appearance of the warp factor has several implications. On the one
hand, warp factors have been argued
\cite{Randall:1999ee,hepth0105097,DWG} to be able to generate
exponential hierarchies between the effective energy scales at
different locations on the internal space, such that gauge fields
localized on different D-branes may experience suppression or
enhancement of gravitational and gauge-theoretical effects. On the
other hand, the appearance of the warp factor implies that our
actual starting point, a direct product $\mathbb{R}^4 \times
\mathbb{T}^6$, no longer solves the ten-dimensional equations of
motion. Thus, in principle, the warp factor should be taken into
account in a dimensional reduction. However, in the large volume
limit the warp factor scales to a constant away from singular
sources, and it has been argued that one may completely neglect it
in this limit \cite{hepth0105097}. At the classical level, the
overall volume is a free parameter, as is manifest from the
no-scale structure of the effective potential (see
\cite{Cremmer:1983bf,Ellis:1983sf} for its phenomenological implications). Thus one can choose an
arbitrarily large value and consider the direct-product ansatz as
an approximate solution. However, as soon as quantum corrections
to the effective four-dimensional action are taken into account,
this line of reasoning should be modified, as they seem to spoil
the no-scale structure \cite{Becker:2002nn,Kachru:2003aw}. The
same is true if one modifies the model by including
higher-dimensional branes
\cite{Blumenhagen:2003vr,Cascales:2003zp}.
These are in principle capable of fixing the overall volume, but in any attempt to do so one
has to make sure that this is done at a value sufficiently large that one can still neglect the warp factor, if the
four-dimensional effective action derived via dimensional reduction is to remain reliable.
In this context we also include some comments on the perspectives of fixing the overall volume by adding
world-volume gauge fluxes on calibrated D9-branes. It appears that
a fixing of the volume at the string scale would be difficult to avoid, which
could invalidate the effective approach. \\

The paper is organized as follows.
In section \ref{sec3} we develop a systematic procedure to apply T-duality to
type I string theory on the level of the effective Lagrangian including the coupling
to gauge fields, and show how to add modifications due to 3-form fluxes in the
background. In section
\ref{sec4} we discuss the effective potential that arises through
this procedure in some detail, its relation to the formulas known from
gauged supergravity, and the truncation to ${\cal N}=1$ supersymmetry.
In section
\ref{sec5} the equations of motions are derived and various forms of constraints
produced. We also discuss briefly how the four-dimensional effective action of the previous
sections can be justified by the large volume scaling argument. In section
\ref{sec6} we finally
add a couple of comments on the addition of higher-dimensional D-branes subject
to certain calibration conditions. The appendix collects some technical
material and additions to the main body of the paper. \\

Before getting started, let us mention the following caveat. Unfortunately,
there is no standard definition of the Hodge-star in the literature about
the class of models under consideration. Our definition, given in appendix \ref{conventions},
differs from the one used e.g.\ in \cite{hepth0009211,hepth0105097,hepth0201028} but coincides with the
one used in \cite{hepth0206241,hepth0207135,hepth0211027,D'Auria:2003jk}. Thus,
what is called an imaginary self-dual (ISD) 3-form flux in \cite{hepth0009211,hepth0105097,hepth0201028}
is imaginary anti-self-dual (IASD) in our conventions and vice versa.


\section{Construction of the action via T-duality}
\label{sec3}

In this section we perform six T-dualities of type I theory.
This operation defines the effective action of type I$'$ on the
dual six-torus. The main objective, then, is to find the modifications in the
action when complex 3-form fluxes are added to the background, following the philosophy outlined
in the introduction. We first go through
the procedure ignoring the coupling to the non-abelian vector fields
of the open string sector, and only then reconsider the full model. \\

Our starting point is type I supergravity, which, including the coupling to non-abelian
vectors of the gauge group $SO(32)$, is described by the action
\beqn \label{iaction}
{\cal S}_{\rm I} & = &
\frac{1}{2 \kappa_{10}^2} \int d^{10} x\, \sqrt{-g} \Big(
  e^{-2 \Phi} \left(R
  + 4 \partial_\mu \Phi \partial^\mu \Phi \right) - \frac12 |\tilde F_3|^2
\Big) \non
&&
- \frac{1}{2 g_{10}^2} \int d^{10} x\, \sqrt{-g}
  e^{-\Phi} {\rm tr}\, |{\cal F}|^2\ ,
\eeqn
where ${\cal F} = \frac12 {\cal F}^a_{MN} T^a dx^M \wedge dx^N$ is given by
\be \label{fmn}
{\cal F}^a_{MN} = 2 \partial_{[M} A^a_{N]} + f^{abc} A^b_M A^c_N
\ee
and we use the definition
\be \label{ftilde}
\tilde F_3 = d C_2 - \frac{\kappa^2_{10}}{g_{10}^2} \omega_3
\ee
with $\omega_3$ the gauge CS 3-form
\be \label{omega3}
\omega_3 = {\rm tr} \left( A \wedge d A
-i \frac23  A \wedge A \wedge A \right)\ .
\ee
In general one defines the Chern Simons forms
$\omega_{2j - 1}$ by
\be \label{csform}
d \omega_{2j - 1} = {\rm tr}\, {\cal F}^j\ .
\ee
Except for all the terms involving the vector fields,
the type I action (\ref{iaction}) is obtained by
quotienting world sheet parity $\Omega$ out of the type IIB action\footnote{The
issue of the self-duality constraint on $F_5$ will be dealt with later.}
\beqn \label{iib}
{\cal S}_{\rm IIB} & = & \frac{1}{2 \kappa_{10}^2}
\int d^{10} x\, \sqrt{-g} \left(
e^{-2 \Phi} \left(R + 4 \partial_\mu \Phi \partial^\mu \Phi
    - \frac12 |H_3|^2 \right) \right. \\
&& \quad\quad \left.
- \frac12 \left( |F_1|^2 + |F_3|^2
                 + \frac12 |F_5|^2 \right) \right)
- \frac{1}{4 \kappa_{10}^2}
  \int C_4 \wedge dB_2 \wedge dC_2 \nonumber\ ,
\eeqn
with
\beqn \label{rrfs}
F_3 & = & dC_2 + C_0 dB_2\ , \non
F_5 & = & dC_4 +
  \frac12 C_2 \wedge dB_2 - \frac12 B_2 \wedge dC_2
\eeqn
and $H_3 = dB_2,\ F_1 = dC_0$.
After modding out $\Omega$, only the RR 2-form enters (\ref{iaction}) via
$F_3 = dC_2$, while the NSNS 2-form $B_2$ is projected out.
Similarly, the T-dual version is a truncation of
the type IIB theory (modding out the T-dual $\Omega$-projection (\ref{modwsp}))
coupled to vectors.
The duality operation then replaces the degrees of freedom $\{ g_{IJ}, C_2, \Phi \}$
of type I by those of type I$'$, which are
\beqn \label{spectrum}
1\, \, {\rm graviton} & : & g_{\mu \nu}\ , \non
12\, \, {\rm gauge\, \, bosons} & : & B_{i\mu}\ ,\  C_{i\mu}\ , \non
38\, \, {\rm scalars} & : & g_{ij}\ ,\  C_{ijkl}\ ,\  \tau = C_0 + i e^\Phi\ .
\eeqn
In terms of ${\cal N}=4$ supersymmetry, these make up a spin-2 and
six abelian vector multiplets. In addition, there are vector
multiplets with bosonic field content $(A^a_\mu, A^a_i)$ in the
adjoint of the gauge group, a subgroup of $SO(32)$. These fields
are referred to as open string fields and arise from the presence
of D3-branes, the T-dual images of the D9-branes of type I. The
O9-plane of type I theory splits into 64 O3-planes, but in the
presence of fluxes the number $N_{\rm D3}$ of D3-branes needed to
cancel their charge is no longer fixed. Rather, turning on 3-form
flux modifies the tadpole condition to \cite{hepth0105097}
\be \label{modtad}
\frac12 N_{\rm flux} + N_{\rm D3} = 16\ ,
\ee
thus effectively replacing some of the D3-branes by flux. The precise form
of $N_{\rm flux}$ will be given later, cf.\ (\ref{nflux}). \\
%


\subsection{T-duality rules}
\label{rules}

For some of the conventions used in the following we refer the
reader to the appendix. Using the standard Kaluza-Klein (KK) ansatz for the metric
\be \label{metricansatz}
ds_{10}^2 = g_{IJ} dx^I dx^J
 = g_{\mu\nu} dx^\mu dx^\nu +
   G_{ij} ( dy^i + A^i_\mu dx^\mu ) ( dy^j + A^j_\nu dx^\nu )\ ,
\ee
the transformations of the NSNS sector can be deduced from e.g.\ \cite{hepth9401139}.
The formulas for the replacements according to dualizing all six
circles simplify for type I to
\be \label{tdmetric}
G_{ij} \mapsto G^{ij}\ , \qquad
g_{\mu \nu} \mapsto g_{\mu \nu}
\ , \qquad
\ee
and
\be \label{tdB}
g_{\mu k} \mapsto G^{ki} B_{\mu i} .\qquad
\ee
This last operation (\ref{tdB}) amounts
to replacing the KK vectors by $A^i_\mu \mapsto B_{\mu i}$,
since $g_{\mu k} = G_{ik}A^i_\mu$.
Finally, the dilaton transforms according to
\be \label{tddilaton}
e^{2 \Phi} \mapsto \frac{e^{2 \Phi}}{{\rm det}(G_{ij})}\ .
\ee
Note that we do not
distinguish the fields of type I or its T-dual from their ancestor fields in
type IIB, since we identify the effective action that is
obtained by projection from type IIB with that of type I (respectively its T-dual)
when the open string vector fields are set to zero. \\

The transformation properties of the RR fields on a
higher-dimensional torus can be found e.g.\ in
\cite{hepth9907132,hepth9912236,hepth0008149}. We will stick to the
formalism of \cite{hepth9907132} here, which leads
to the same results as the one of \cite{hepth9912236,hepth0008149}.
It was shown there that the type IIB equations of motion for the action
(\ref{iib}) can alternatively be derived from an action that is manifestly
invariant under T-duality, where the second line of (\ref{iib}), the
kinetic terms for the RR forms plus the CS term, is replaced by\footnote{A similar democratic version
of type II supergravity was discussed in \cite{BKORVP}.}
\be \label{iib2}
{\cal S}_{\rm RR+CS} ~\rightarrow~
- \frac{1}{8 \kappa_{10}^2 } \sum_{q=1}^5 |F_{2q-1}|^2\ ,
\ee
and $F_p$ is now defined more generally as
\be \label{genrrforms}
\sum_{q=1}^5 F_{2q-1} = e^{-B_2} \wedge \sum_{q=0}^4 dD_{2q}\ .
\ee
It is important that no CS term needs to be included explicitly in (\ref{iib2}).
It would appear automatically if one dualizes the forms of high degree, see \cite{hepth9907132}
for more details.
The $D_p$ transform in the spinor representation of $O(6,6,\mathds{Z})$
\cite{hepth9907132,hepth9912236,hepth0008149}. To complete the definitions note
\beqn \label{CD}
D_0 = C_0\ , \quad
D_2 = C_2 + C_0 B_2\ , \quad
D_4 = C_4 + \frac12 B_2 \wedge C_2 + \frac12 C_0 B_2^2\ ,
\eeqn
and
\be \label{defhodge}
* F_1 = F_9\ , \quad
* F_3 = -F_7\ , \quad
* F_5 = F_5\ ,
\ee
the latter to be imposed after deriving the equations of motions
from the action. One can actually give a more explicit definition
for the field strengths of higher degree and we will
use this later in (\ref{generalf}), see e.g.\ \cite{BKORVP}. \\

We are now interested only in the particular element of the
whole $O(6,6,\mathds{Z})$ symmetry group that corresponds to
(\ref{tdmetric}). This element is
given by\footnote{Note that the overall
sign is arbitrary. The derivation of the transformation
rules from the rules given in \cite{hepth9907132} is performed in the appendix
(but see also \cite{hepth0008149}).}
\be \label{tdrr2}
D_{\mu_1 \ldots \mu_q i_1 \ldots i_n} ~\mapsto~  \frac{(-1)^{n(n-1)/2}}{(6-n)!}
\hat \epsilon_{i_1 \ldots i_n}\, ^{i_{n+1} \ldots i_6}
D_{\mu_1 \ldots \mu_q i_{n+1} \ldots i_6}\ ,
\ee
where the hat on $\hat \epsilon$ means that here the
epsilon-symbol takes values
$\hat\epsilon^{1\, ... \, 6} = \hat\epsilon_{1\, ... \, 6} = 1$
and the indices are contracted with Kronecker-deltas instead of the
metric. To write (\ref{tdrr2}) in terms of tensor quantities, one has to
factor out the determinant $G= {\rm det}(G_{ij})$ from the internal
metric to cancel the factor $G$ appearing in
the contractions, that is
\be \label{tdrr3}
D_{\mu_1 \ldots \mu_q i_1 \ldots i_n} ~\mapsto~ \frac{(-1)^{n(n-1)/2} \sqrt{G}}{(6-n)!}
\epsilon^{i_1 \ldots i_6}
D_{\mu_1 \ldots \mu_q i_{n+1} \ldots i_6}\ .
\ee
Now $\epsilon_{1\, ... \, 6} = \sqrt{G},\
\epsilon^{1\, ... \, 6} = 1/\sqrt{G}$ and indices are contracted with
$G_{ij}$.
The field strengths $F_p$ can be transformed
in a similar vein, by simply replacing $D_p$ with $dD_p$ in
(\ref{tdrr3}) \cite{hepth9912236}.
Thus the components of the RR 3-form field strength
have the following transformation properties\footnote{Once more, we would
like to refer to the appendix for the definitions of the notations used here. The upper
indices $\{p,q\}$ refer to the bi-degree as forms on the non-compact and internal part of the
ten-dimensional space-time.}
\beqn \label{c2trans}
(dC_2)^{\{0,3\}}_{ijk} & \mapsto & -
\frac{\sqrt{G}}{3!} \epsilon^{ijklmn} (dD_2)^{\{0,3\}}_{lmn}\ , \non
(dC_2)^{\{1,2\}}_{\mu ij} & \mapsto & -
\frac{\sqrt{G}}{4!} \epsilon^{ijklmn} (dD_4)^{\{1,4\}}_{\mu klmn}\ , \non
(dC_2)^{\{2,1\}}_{\mu \nu i} & \mapsto &
     +\frac{\sqrt{G}}{5!} \epsilon^{ijklmn} (dD_6)^{\{2,5\}}_{\mu  \nu jklmn}\ , \\
(dC_2)^{\{3,0\}}_{\mu \nu \rho} & \mapsto &
   + \frac{\sqrt{G}}{6!} \epsilon^{ijklmn} (dD_8)^{\{3,6\}}_{\mu \nu \rho ijklmn}\ .
\nonumber
\eeqn
It is evident that the T-dual of the type I RR 2-form
also includes RR forms of degree six and eight, which we later dualize
to forms of lower degree.
Finally, according to \cite{hepth9910053} the internal components of the non-abelian
vectors are mapped via
\be \label{mapvect}
A^a_i ~\mapsto~ A^{ai}\ ,
\ee
which is easily applied to ${\cal F}$ and $\omega_3$.
It is important to note that the $A^{ai}$ are actually the independent dual open string moduli,
in other words, the dual scalars $A^a_i$ depend implicitly on the metric. The same is then true
e.g.\ for ${\cal F}_{ij}$ and $\omega_{ijk}$.


\subsection{T-duality without vectors}
\label{tdualwo}

We now apply the rules assembled in the previous section, first to the pure
supergravity part of the type I bosonic action.
Our starting point is (\ref{iaction}) with the vectors erased,
\be \label{typi}
{\cal S}_{\rm I}[A=0] =
\frac{1}{2 \kappa_{10}^2} \int d^{10} x\, \sqrt{-g} \left(
  e^{-2 \Phi} \left(R
  + 4 \partial_\mu \Phi \partial^\mu \Phi \right) - \frac12 |F_3|^2
\right) \ ,
\ee
which we split into
\be \label{splits}
{\cal S}_{\rm I}[A=0] = {\cal S}_{\rm NSNS} + {\cal S}_{\rm RR}\ .
\ee
Let us consider the NSNS part first. Using the rules of the last section
and the usual form of the NSNS action on a torus (see e.g.\
(2.17) of \cite{MaSc}), it is straightforward to verify that it is mapped as
\be \label{snsns}
{\cal S}_{\rm NSNS} \mapsto
\frac{1}{2 \kappa_{10}^2} \int d^{10} x\, \sqrt{-g}
  e^{-2 \Phi} \left(R
  + 4 \partial_\mu \Phi \partial^\mu \Phi
  - \frac14 G^{ij} H_{i \mu \nu} H_j^{\mu \nu} \right)\ ,
\ee
where we defined
\be \label{fkk}
H_{j \nu \sigma} = 2 \partial_{[\nu} B_{\sigma] j}\ .
\ee
Note that after T-duality the metric in (\ref{spectrum}) has no off-diagonal components and that
the original Kaluza-Klein vectors have been mapped to components of the NSNS $B$-field (\ref{fkk}). \\

Let us now consider the RR part.
There is a helpful trick to make the appearance of the KK vectors
explicit \cite{hepth9907132}. Using (\ref{metricansatz})
one can rewrite the kinetic terms for the RR forms as
\be \label{rewriterrforms}
|F_p|^2 = |F'_p|^2_{G_{ij}, g_{\mu\nu}}
\ee
by redefining
\be \label{rr'}
F_p' = F_p|_{dy^i \mapsto dy^i - A^i_\mu dx^\mu}\ .
\ee
On the left-hand side of (\ref{rewriterrforms}) the contractions are performed
with the full metric, whereas on the right-hand side, the off-diagonal part is omitted
and absorbed into a redefinition of the field strength.
In other words, defining new forms as in (\ref{rr'}) one can perform
all contractions using the internal or external components of the metric
only.
In the dual theory $F'_p = F_p$ and we can
omit the prime. Using (\ref{c2trans}), we obtain
\be \label{warumhattedasnochkeinlabel}
(F'_3)^{\{n,3-n\}}_{\mu_1 \ldots \mu_n i_1 \ldots i_{3-n}} ~\mapsto~
(-1)^{(3-n)(2-n)/2} \frac{\sqrt{G}}{(3+n)!} \epsilon^{i_1 \ldots i_6}
(F_{3+2n})^{\{n,3+n\}}_{\mu_1 \ldots \mu_n i_{4-n} \ldots i_{6}}\ .
\ee
The right-hand side exactly takes the form
of (\ref{genrrforms}) subject to the projection in the T-dual model.
This is because the KK vectors of the metric, that appear in the definition
(\ref{rr'}) of $F'_3$, are mapped to those of the NSNS $B$-field exactly in
a way required by  (\ref{genrrforms}).
Applying (\ref{warumhattedasnochkeinlabel})
in (\ref{typi}) maps the RR kinetic term according to
\beqn \label{tdkin}
-\frac{1}{4\kappa^2_{10}} \int d^{10}x \sqrt{- g} |F_3|^2 &
\mapsto &
\frac{1}{4\kappa^2_{10}} \int \left(
F_9^{\{3,6\}} \wedge * F_9^{\{3,6\}} + F_7^{\{2,5\}} \wedge * F_7^{\{2,5\}}
\right. \\
&& \hspace{2cm}
\left.
+ F_5^{\{1,4\}} \wedge * F_5^{\{1,4\}}
+ F_3^{\{0,3\}} \wedge * F_3^{\{0,3\}} \right)\ , \nonumber
\eeqn
where the Hodge star $*$ is with respect to the
ten-dimensional T-dual metric. In applying the T-duality rules of section \ref{rules}
along all six internal directions, we have to assume that none of
the fields depend on the internal coordinates.
This is in particular true for the metric components (\ref{metricansatz}).
Thus we do not consider any non-trivial spin connection for the metric
before T-duality. From the results of \cite{hepth9912236,DaHu,GLMW,KSTT}
this implies that we are not able to produce any non-trivial NSNS flux directly via
T-duality and in (\ref{tdkin}) we implicitly assume $(dB_2)^{\{0,3\}} = 0$. \\

In principle (\ref{tdkin}) gives all the terms in the T-dual theory that come from
the RR 3-form field strength of type I, including a purely internal part
$F_3^{\{0,3\}}$. In order to make contact with
the standard form that would be obtained from type IIB by the T-dual projection, which involves only field strengths
of degree five and lower, we next remove the RR forms of unconventionally high degree
from the action in favor of their dual forms. According to the standard
procedure (see e.g.\ \cite{MO})
this would amount to imposing the Bianchi identity for $F_9^{\{3,6\}}$ and
$F_7^{\{2,5\}}$, respectively, via Lagrange multipliers and then integrating out $F_9^{\{3,6\}}$ and
$F_7^{\{2,5\}}$, leaving the Lagrange multipliers as the dual degrees of freedom.
However, in the presence of 3-form flux $F_3^{\{0,3\}}$
this method does not seem to be applicable in a straightforward 
way.\footnote{The problem derives from the fact that if one naively imposes the
Bianchi identities of $F_9^{\{3,6\}}$ and $F_7^{\{2,5\}}$ one generates terms 
involving different components of the RR field strengths as appearing in (\ref{tdkin}), like 
e.g.\ $F_7^{\{4,3\}}$. Their proper treatment requires one to use a democratic version of type I
supergravity, involving $C_2$ and $C_6$ on the same footing before T-duality, cf.\ appendix 
\ref{sdactionsec}. After T-duality this would also include kinetic terms for $F_7^{\{4,3\}}$ etc.,
which make it possible to properly dualize these components. \label{demoaction}}
We therefore
follow a different strategy, first setting also the RR 3-form flux in (\ref{tdkin}) to zero.
For this case we perform the dualization of $F_9^{\{3,6\}}$ and $F_7^{\{2,5\}}$ and only afterwards infer the
effects of non-vanishing flux in the effective action.
\\

So let us for the moment consider (\ref{tdkin}) without the 3-form flux.
In order to dualize $F_9^{\{3,6\}}$ we impose its
Bianchi identity by adding a Lagrange multiplier term
\be \label{lagrm}
\delta {\cal S} = -\frac{1}{2\kappa^2_{10}}
\int C_0 d^{\{1,0\}} \Big( F_9^{\{3,6\}} + B_2^{\{1,1\}} \wedge (dD_6)^{\{2,5\}} - \frac12 (B_2^{\{1,1\}})^2 \wedge
(dD_4)^{\{1,4\}} \Big)\ .
\ee
We called the Lagrange multiplier $C_0$, anticipating that it will be identified
with the RR scalar in a moment. This becomes clear by inspection of its kinetic term
(\ref{kinc0}) below and by comparison with the (truncated) type IIB action.
A partial integration in the first term of (\ref{lagrm}) leads to
\be \label{lagrm2}
\delta {\cal S} = \frac{1}{2\kappa^2_{10}}
\int (dC_0)^{\{1,0\}} \wedge F_9^{\{3,6\}}
- \frac{1}{2\kappa^2_{10}} \int C_0\, (dB_2)^{\{2,1\}} \wedge F_7^{\{2,5\}}  \ .
\ee
Now integrating out $F_9^{\{3,6\}}$ through
\be \label{intout}
* F_9^{\{3,6\}} = (dC_0)^{\{1,0\}}
\ee
and plugging this back into the action produces a kinetic term for $C_0$:
\be \label{kinc0}
\frac{1}{4\kappa^2_{10}} \int (dC_0)^{\{1,0\}}  \wedge * (dC_0)^{\{1,0\}} =
- \frac{1}{4\kappa_{10}^2} \int d^{10}x\, \sqrt{-g} |F_1^{\{1,0\}}|^2 \ .
\ee
If we next want to integrate out $F_7^{\{2,5\}}$ we impose its
Bianchi identity by adding a Lagrange multiplier term
\be \label{lagrm3}
\delta {\cal S} = \frac{1}{2\kappa^2_{10}}
\int C_2^{\{1,1\}} \wedge d^{\{1,0\}} \Big( F_7^{\{2,5\}} + B_2^{\{1,1\}} \wedge dD_4^{\{1,4\}} \Big)\ .
\ee
Again we anticipate that the Lagrange multiplier is identified with the RR 2-form,
as can be read off from (\ref{kinc2}) below. Then it is also clear that the only relevant
component of $C_2$ is $C_2^{\{1,1\}}$. Performing a partial integration for the
first term of (\ref{lagrm3}) leads to
\be \label{partint}
\delta {\cal S} = - \frac{1}{2\kappa^2_{10}} \int (dC_2)^{\{2,1\}} \wedge F_7^{\{2,5\}} +
\frac{1}{2\kappa^2_{10}} \int (C_2)^{\{1,1\}} \wedge
                        (dB_2)^{\{2,1\}} \wedge (dC_4)^{\{1,4\}} \ .
\ee
%
Together with (\ref{lagrm2}), integrating out $F_7^{\{2,5\}}$ now gives
\be \label{intout2}
* F_7^{\{2,5\}} = - (dC_2)^{\{2,1\}} - C_0 (dB_2)^{\{2,1\}} \ .
\ee
Inserting this into the action we obtain the kinetic term
\beqn \label{kinc2}
\frac{1}{4\kappa^2_{10}} \int \left((dC_2)^{\{2,1\}} +  C_0 (dB_2)^{\{2,1\}}\right) \wedge *
\left((dC_2)^{\{2,1\}} +  C_0 (dB_2)^{\{2,1\}}\right)  &&
\\
&& \hspace{-5cm} =~  - \frac{1}{4\kappa^2_{10}} \int  d^{10}x\, \sqrt{-g} |F_3^{\{2,1\}}|^2\ ,
\nonumber
\eeqn
and a Chern-Simons term
\be \label{cs}
{\cal S}_{\rm CS} =
\frac{1}{2\kappa^2_{10}} \int (dC_2)^{\{1,1\}} \wedge (dB_2)^{\{2,1\}}
\wedge C_4^{\{0,4\}}\ .
\ee
In this way, rewriting the T-dual of the type I action in terms of
potentials with degree up to four produces the
correct CS term that one expects from truncating type IIB,
even though type I does not possess such a term on its own.
In all, this action is exactly what one would get from a reduction of
type IIB subject to imposing the self-duality constraint on $F_5$,
as we verify in appendix \ref{kinc4}. \\

Let us now study what changes should occur due to inclusion of
3-form flux. As was explained in the introduction, we make use of the fact that
type I$'$ theory with all non-abelian vector fields set to zero
is a truncation of type IIB by modding out the T-dual projection (\ref{modwsp}).
It is obvious from the last term of (\ref{tdkin}) that a scalar potential
appears in the presence of flux. From (\ref{tdkin}) we would infer a
term\footnote{Calling this term a potential is slightly imprecise,
as (\ref{pote}) is still a term in the T-dual ten-dimensional action.
What we mean is of course that this term leads to a potential
in the effective theory - after dimensional reduction.}
\be \label{pote}
- \frac{1}{4\kappa^2_{10}}
\int d^{10}x \sqrt{-g} |F_3^{\{0,3\}}|^2 \ ,
\ee
where, as mentioned below (\ref{tdkin}), only a
$F_3^{\{0,3\}} = (dC_2)^{\{0,3\}}$ term arises directly from T-duality.
However, comparing to the type IIB action truncated by (\ref{modwsp}),
the expression (\ref{pote}) for the potential receives an additional
contribution from the NSNS sector of the type IIB action, i.e.\ from the term
\be
- \frac{1}{4 \kappa^2_{10}} \int d^{10}x \sqrt{-g} e^{-2 \Phi}
|H_3^{\{0,3\}}|^2\ ,
\ee
and another modification due to the fact that the RR $3$-form field strength actually appears in
the combination (\ref{rrfs}). Then
the total potential can be expressed via the complex combination
\be \label{G}
G_3 = dC_2 + \tau dB_2 = F_3 + i e^{-\Phi} H_3 \ ,
\ee
where $\tau = C_0 + i e^{-\Phi}$, in the form
\be \label{pote2}
{\cal S}_{\rm pot}
= - \frac{1}{4\kappa^2_{10}} \int d^{10}x \sqrt{-g} |G_3^{\{0,3\}}|^2\ .
\ee
Coupling the theory to the D3-branes will, besides other modifications,
give rise to further contributions to the potential, coming from the
world-volume scalars and from the tension of the branes. From the point of view
of the truncated type IIB theory, (\ref{pote2}) is the obvious part of the potential,
while from the point of view of type I$'$ it is the other way around and (\ref{pote2})
cannot be derived directly via T-duality.
Since the internal components $C^{\{ 0,2\}}_2$ and $B^{\{ 0,2\}}_2$ are projected out of the
spectrum, $G^{\{0,3\}}_3$ is not a flux for any (dynamical) field strength in type I$'$, but just some antisymmetric
background parameter. It is then not obvious that a potential term (\ref{pote2})
can appear as part of a consistent modification of the Lagrangian that allows
supersymmetry to be preserved on the level of the action.
It can be deduced from a comparison with type IIB,
but the systematic approach to determine
such potentials is through gauged supergravity \cite{hepth0206241,hepth0211027,D'Auria:2003jk}. \\

Furthermore, the kinetic term for the
scalars $C_4^{\{0,4\}}$,
i.e.\ the penultimate term in (\ref{tdkin}), is modified in the presence of fluxes.
Again, via T-duality one can only infer a correction
$F_5^{\{1,4\}} = (dC_4)^{\{1,4\}} - \frac12 B_2^{\{1,1\}} \wedge
(dC_2)^{\{0,3\}}$, but comparison with type IIB (\ref{rrfs}) shows that
actually the combination
\be \label{dc4mod}
F_5^{\{1,4\}} = (dC_4)^{\{1,4\}} - \frac12 B_2^{\{1,1\}} \wedge
(dC_2)^{\{0,3\}} + \frac12 C_2^{\{1,1\}} \wedge (dB_2)^{\{0,3\}}
\ee
appears.
Finally, when dualizing $F_9^{\{3,6\}}$ and $F_7^{\{2,5\}}$ one would in principle have to adapt
their Bianchi identities in (\ref{lagrm}) and
(\ref{lagrm3}). Fortunately, this would just influence the resulting
Chern-Simons term and not the kinetic terms for $C_0$,
$C_2^{\{1,1\}}$ and $B_2^{\{1,1\}}$. Hence, the kinetic terms in the RR sector after
T-duality are
\beqn \label{kinrr}
{\cal S}_{\rm kin} &=& -
\frac{1}{4\kappa^2_{10}} \int d^{10}x \sqrt{-g} \Big(
|(dC_0)^{\{1,0\}}|^2 + |(dC_2)^{\{2,1\}} + C_0 (dB_2)^{\{2,1\}}|^2 \non
&& \hspace{2cm} +~ |(dC_4)^{\{1,4\}} - \frac12 B_2^{\{1,1\}} \wedge
(dC_2)^{\{0,3\}} + \frac12 C_2^{\{1,1\}} \wedge (dB_2)^{\{0,3\}}|^2
\Big) \non &=& - \frac{1}{4\kappa^2_{10}} \int d^{10}x \sqrt{-g}
\left( |F_1^{\{1,0\}}|^2 + |F_3^{\{2,1\}}|^2 + |F_5^{\{1,4\}}|^2 \right)\ .
\eeqn
We see that under a
gauge transformation of the Kaluza-Klein vectors $C_2^{\{1,1\}}$ and
$B_2^{\{1,1\}}$, the scalars $C_4^{\{0,4\}}$ have to transform
in order to render their kinetic term gauge
invariant. This complication can be turned to our advantage in that
it reveals the necessary modification of the CS terms as in \cite{TsZi}.
The Chern-Simons term must be modified to
\beqn \label{csmod}
2\kappa^2_{10} {\cal S}_{\rm CS} & = &
\int (dC_2)^{\{2,1\}} \wedge (dB_2)^{\{2,1\}} \wedge C_4^{\{0,4\}} \non
&& -~ \frac12 \int C_2^{\{1,1\}} \wedge (dB_2)^{\{2,1\}}
\wedge B_2^{\{1,1\}} \wedge (dC_2)^{\{0,3\}} \non
&& -~ \frac12 \int B_2^{\{1,1\}} \wedge (dC_2)^{\{2,1\}}
\wedge C_2^{\{1,1\}} \wedge (dB_2)^{\{0,3\}}\ .
\eeqn
In summary, the RR part of the action (without vectors) is of the form
\be \label{act}
{\cal S}_{\rm RR} = {\cal S}_{\rm pot} + {\cal S}_{\rm kin} + {\cal S}_{\rm CS}\ ,
\ee
where  ${\cal S}_{\rm pot}$, ${\cal S}_{\rm kin}$ and ${\cal S}_{\rm CS}$
are given by (\ref{pote2}), (\ref{kinrr}) and (\ref{csmod}). \\

To illustrate the notation let us write out the covariant
derivative for the axions descending from $C_4$. Since the only
dynamical components of $B_2$ and $C_2$ are $B_2^{\{1,1\}}$ and
$C_2^{\{1,1\}}$, the last term in (\ref{kinrr}) is proportional to
the square of
\be \label{covder}
\partial_\mu (C_4)_{ijkl}
- 2 (B_2)_{\mu [i} (dC_2)_{jkl]}
+ 2 (C_2)_{\mu [i} (dB_2)_{jkl]} \ .
\ee
In order to make contact with the standard conventions in supergravity,
as used in \cite{hepth0201029,hepth0206241,D'Auria:2003jk} in particular, let us introduce a different
parameterization of the axionic scalars
\beqn \label{repara}
\partial_\mu (C_4)_{ijkl} &=&
    \frac{1}{2} \frac{1}{\sqrt{G}} \epsilon_{ijklmn} \partial_\mu \beta^{mn} \ , \non
(B_2)_{\mu [i} (dC_2)_{jkl]}  &=&
  - \frac{1}{8} \epsilon_{ijklmn} ( \star dC_2 )^{mnp} (B_2)_{\mu p}\ ,
\eeqn
and similarly for the third term,
with $\star$ denoting the six-dimensional internal Hodge operator.
This leads to a covariant derivative
\be \label{covderbeta}
D_\mu \beta^{ij} = \partial_\mu \beta^{ij}
+ \frac12 \sqrt{G} (\star dC_2)^{ijk} (B_2)_{\mu k}
- \frac12 \sqrt{G} (\star dB_2)^{ijk} (C_2)_{\mu k}
\ee
for $\beta^{ij}$. Note that the combinations $\sqrt{G} (\star dC_2)^{ijk}$ and
$\sqrt{G} (\star dB_2)^{ijk}$ are actually constant, independent of the metric,
for constant background fluxes $(dC_2)_{ijk}$ and $(dB_2)_{ijk}$. They correspond to
the flux parameters $f^{\Lambda \Sigma \Gamma}_\alpha$ of \cite{hepth0206241}, while the $\beta^{ij}$
serve as axionic moduli also independent of the metric.
Later we will come back to the precise relation between our notation and the
one used in \cite{hepth0206241,D'Auria:2003jk}.
In the light of (\ref{c2trans}), the reparameterization (\ref{repara})
just reintroduces the T-dual variables, which reflects the fact that
the dependence of mass parameters on the radii of the torus was found to be
inverted in \cite{hepth0206241}.


\subsection{Coupling to vector fields}

In the previous section we have performed the T-duality of the type I action
without vector fields, reorganized the RR forms to be able to compare to the
truncated type IIB action, and then deduced the modification due to 3-form
fluxes. We
now want to add in the CS correction appearing in (\ref{ftilde}) and
the Yang-Mills action for the vectors. Thus, our
starting point is now (\ref{iaction}).
In order to T-dualize (\ref{iaction})
we need to make the appearance of the KK vectors
in the terms involving ${\cal F}$ and $\omega_3$
explicit. Let us take ${\cal F}$ first.
On a torus we have
\beqn \label{fmnexplicit}
{\cal F}^a_{\mu\nu} &=& 2 \partial_{[\mu} A^a_{\nu]} + f^{abc}
  A_\mu^b A_\nu^c \ , \non
{\cal F}^a_{\mu i} &=& D_\mu A^a_i ~=~ \partial_\mu A^a_i + f^{abc}
  A_\mu^b A_i^c \ , \non
{\cal F}^a_{ij} &=& f^{abc} A_i^b A_j^c \ .
\eeqn
Our notation does not distinguish scalars $A_i$ from vector fields
$A_\mu$ and we do not use a background gauge flux $f_{ij}^a$. This
is related to the fact that we intend to stick to D3-branes after
the T-duality. We shall come back to this point later. As in
(\ref{rr'}) we introduce the components of ${\cal F'}$ as
\beqn \label{f'}
{\cal F'}^a_{\mu \nu} & = & {\cal F}^a_{\mu \nu} + 2 A^i_{[\mu} {\cal F}^a_{\nu] i} +
{\cal F}^a_{ij} A^i_{[\mu} A^j_{\nu]}\ , \non
{\cal F'}^a_{\mu i} & = & {\cal F}^a_{\mu i} + {\cal F}^a_{ij}  A^j_{\mu}\ ,  \\
{\cal F'}^a_{ij} & = & {\cal F}^a_{ij}\ , \nonumber
\eeqn
where ${\cal F}^a_{MN}$ is defined as in (\ref{fmn}). Using new
vector fields, invariant under KK gauge transformations,
\be \label{newvect}
\tilde A^a_\mu = A^a_\mu - A^a_i A^i_\mu\ ,
\ee
it is straightforward to verify the relations
\beqn \label{tildef'}
{\cal F'}^a_{\mu i} & = & \tilde D_\mu A^a_i\ , \\
{\cal F'}^a_{\mu \nu} & = & \tilde {\cal F}^a_{\mu \nu} + 2 \partial_{[\mu} A^i_{\nu]}\, A^a_i\ , \nonumber
\eeqn
where we used
\beqn \label{def}
\tilde D_\mu A^a_i & = & \partial_\mu A^a_i + f^{abc} \tilde A^b_\mu A^c_i \ , \non
\tilde {\cal F}^a_{\mu \nu} & = &  2 \partial_{[\mu} \tilde A^a_{\nu]}
+ f^{abc} \tilde A^b_\mu \tilde A^c_\nu\ .
\eeqn
Thus, using (\ref{mapvect}), the kinetic term for the vectors is
mapped as follows under T-duality:
\beqn \label{tdym}
\int d^{10} x\, \sqrt{-g}
  e^{-\Phi} {\rm tr}\, |{\cal F}|^2 & \mapsto &
\int d^{10} x\, \sqrt{-g_4} e^{-\Phi} \Big( G_{ij} g^{\mu \nu}
  \tilde D_\mu A^{ai} \tilde D_\nu A^{aj} \non
&& +~ \frac12 g^{\mu \nu} g^{\rho \sigma} (\tilde {\cal F}^a_{\mu \rho} + H_{i\mu \rho}A^{ai})
(\tilde {\cal F}^a_{\nu \sigma} + H_{j \nu \sigma}A^{aj})  \\
&& +~ \frac12 G_{ij} G_{kl} f^{abc} f^{ade} A^{bi} A^{ck} A^{dj} A^{el} \Big)\ , \nonumber
\eeqn
where now
\be \label{tilda}
\tilde A^a_\mu = A^a_\mu - A^{ai} B_{\mu i}\ ,
\ee
and $H_{i\mu \nu}$ is given by (\ref{fkk}).
The term in the last line of (\ref{tdym})
represents a contribution to the potential for the
open string fields and can also be written as tr$|{\cal F}^{\{0,2\}}|^2$ up to a constant.
Notice that (\ref{tdym}) is separately invariant under the gauge transformations
\be \label{gaugetrafo1}
B_{\mu i} \rightarrow B_{\mu i} + \partial_\mu \epsilon_i\ ,
\ee
with $\tilde A^a_\mu$ and $A^{ai}$ inert, and
\be \label{gaugetrafo2}
\tilde A^a_\mu \rightarrow \tilde A^a_\mu + \partial_\mu \epsilon^a + f^{abc} \tilde A^b_\mu \epsilon^c \quad ,
\quad A^{ai} \rightarrow A^{ai} + f^{abc} A^{bi} \epsilon^c\ ,
\ee
with $B_{\mu i}$ inert. \\

Let us now turn to the mapping of $\omega_3$ under the duality
transformation, shifting to $\omega_3'$ as before. It is given by
\beqn \label{omtrans}
(\omega_3')^{\{0,3\}}_{ijk} &\mapsto&
 \frac{1}{3!} \epsilon^{ijklmn}
 \left( \star \omega_3 \right)^{\{0,3\}}_{lmn} \ ,
\\
(\omega_3')^{\{1,2\}}_{\mu ij} &\mapsto&
 \epsilon^{ijklmn}
 \left( \frac{1}{4!}
 \left( \star \omega_3 \right)^{\{1,4\}}_{\mu klmn}
 - \frac{1}{3!} (B_2)^{\{1,1\}}_{\mu k} \left( \star \omega_3 \right)^{\{0,3\}}_{lmn}
 \right) \ ,
\non
\frac{1}{2} (\omega_3')^{\{2,1\}}_{\mu\nu i} &\mapsto&
 \epsilon^{ijklmn}
 \left( \frac12 \frac{1}{5!}
 \left( \star \omega_3 \right)^{\{2,5\}}_{\mu\nu jklmn}
\right.
\non
&& \left. \hspace{.8cm}
 - \frac{1}{4!} (B_2)^{\{1,1\}}_{[\mu |j|} \left( \star \omega_3 \right)^{\{1,4\}}_{\nu] klmn}
 + \frac12 \frac{1}{3!} (B_2)^{\{1,1\}}_{[\mu |j|} (B_2)^{\{1,1\}}_{\nu] k}
   \left( \star \omega_3 \right)^{\{0,3\}}_{lmn}
 \right)
\ , \non
\frac{1}{3!} (\omega_3')^{\{3,0\}}_{\mu\nu\rho} &\mapsto&
 \epsilon^{ijklmn}
 \left( \frac{1}{3!} \frac{1}{6!}
 \left( \star \omega_3 \right)^{\{3,6\}}_{\mu\nu\rho ijklmn}
 -\frac12 \frac{1}{5!} (B_2)^{\{1,1\}}_{[\mu |i|} \left( \star \omega_3 \right)^{\{2,5\}}_{\nu\rho] jklmn}
\right.
\non
&& \hspace{.8cm}
 + \frac12 \frac{1}{4!} (B_2)^{\{1,1\}}_{[\mu |i|} (B_2)^{\{1,1\}}_{\nu |j|}
 \left( \star \omega_3 \right)^{\{1,4\}}_{\rho] klmn} \non
&& \left. \hspace{.8cm}
 - \frac{1}{(3!)^2} (B_2)^{\{1,1\}}_{[\mu |i|} (B_2)^{\{1,1\}}_{\nu |j|}  (B_2)^{\{1,1\}}_{\rho] k}
   \left( \star \omega_3 \right)^{\{0,3\}}_{lmn}
 \right) \ . \nonumber
\eeqn
Since $\star$ denotes the six-dimensional internal Hodge operation,
$\star \omega_3$ is a formal sum of forms of degree 3, 5, 7 and 9:
\be
\star \omega_3 = (\star \omega_3)^{\{0,3\}} + (\star \omega_3)^{\{1,4\}}
+ (\star \omega_3)^{\{2,5\}} + (\star \omega_3)^{\{3,6\}} \ .
\ee
Together with (\ref{c2trans}) we can then write
\beqn
( \tilde F_3' )^{\{p,3-p\}}_{\mu_1
\ldots \mu_p i_1 \ldots i_{3-p}} \mapsto
- \frac{(-1)^{p(p-1)/2}\sqrt{G}}{(p+3)!}
\epsilon^{i_1 \ldots i_{3-p}j_1 \ldots j_{3+p}}
(\hat F_{3+2p}^{\{p,p+3\}})_{\mu_1 \ldots \mu_p j_1 \ldots j_{3+p}} \ ,
\eeqn
where we have defined
\be \label{fstr+cs}
\hat F_{3+2p}^{\{p,p+3\}} =
\Big[
e^{-B} \wedge \sum_{q = 0}^p
\Big( dD + \gamma
            (-1)^{q(q-1)/2} \star \omega_3
\Big)^{\{q,q+3\}}
\Big]^{\{ p,p+3 \}} \ .
\ee
with the abbreviation
\be \label{gamma}
\gamma = \frac{\kappa^2_{10}}{g_{10}^2\sqrt{G}}\ .
\ee
Using this rule, (\ref{tdkin}) now becomes
\beqn \label{tdkin2}
-\frac{1}{4\kappa^2_{10}} \int d^{10}x \sqrt{-g} |\tilde F_3|^2 &
\mapsto &
\frac{1}{4\kappa^2_{10}} \int \left( \hat F_9^{\{3,6\}} \wedge * \hat F_9^{\{3,6\}}
+ \hat F_7^{\{2,5\}} \wedge * \hat F_7^{\{2,5\}}
\right. \\
&& \hspace{1.5cm} \left.
+ \hat F_5^{\{1,4\}} \wedge * \hat F_5^{\{1,4\}}
+ \hat F_3^{\{0,3\}} \wedge * \hat F_3^{\{0,3\}} \right)\ . \nonumber
\eeqn
The potential term $|\hat F_3^{\{0,3\}}|^2$ is now modified due to
the open string scalars. For vanishing 3-form flux and taken
together with the potential term of (\ref{tdym}), this is the
potential known also for the
heterotic string \cite{ED1,Ferrara:1986qn,ED}. \\

To eliminate the RR forms of high degree,
we now follow the same procedure as before and assume vanishing
3-form flux while dualizing $\hat F_9^{\{3,6\}}$ and $\hat F_7^{\{2,5\}}$.
In order to dualize $\hat F_9^{\{3,6\}}$, we impose its
Bianchi identity by adding the Lagrange multiplier term (anticipating that
the Lagrange multiplier will be identified with the RR scalar as in (\ref{lagrm}))
\beqn \label{lagrm4}
\delta {\cal S} & = & -\frac{1}{2\kappa^2_{10}}
\int C_0 d^{\{1,0\}} \Big[ \hat F_9^{\{3,6\}} + B_2^{\{1,1\}} \wedge (dD_6)^{\{2,5\}} - \frac12 (B_2^{\{1,1\}})^2 \wedge
(dD_4)^{\{1,4\}} \non
&& \hspace{3cm} + \gamma \Big( \left( \star \omega_3 \right)^{\{3,6\}}
- B_2^{\{1,1\}} \wedge \left( \star \omega_3 \right)^{\{2,5\}} \\
&& \hspace{3cm} - \frac12 (B_2^{\{1,1\}})^2 \wedge \left( \star \omega_3 \right)^{\{1,4\}}
+ \frac{1}{3!} (B_2^{\{1,1\}})^3 \wedge \left( \star \omega_3 \right)^{\{0,3\}} \Big) \Big] \non
& = & \frac{1}{2\kappa^2_{10}} \int (dC_0)^{\{1,0\}} \wedge \hat F_9^{\{3,6\}} - \frac{1}{2\kappa^2_{10}}
\int C_0 \Big[ (dB_2)^{\{2,1\}} \wedge \hat F_7^{\{2,5\}} \non
&& \hspace{.5cm} \mbox{} + d^{\{1,0\}} \left(\gamma (\star \omega_3)^{\{3,6\}}\right)
- B_2^{\{1,1\}} \wedge d^{\{1,0\}} \left(\gamma (\star \omega_3)^{\{2,5\}}\right) \non
&& \hspace{.5cm} \mbox{} - \frac12 (B_2^{\{1,1\}})^2 \wedge d^{\{1,0\}}
\left( \gamma (\star \omega_3)^{\{1,4\}} \right)
+ \frac{1}{3!} (B_2^{\{1,1\}})^3 \wedge d^{\{1,0\}}
\left( \gamma (\star \omega_3)^{\{0,3\}} \right) \Big]\ . \nonumber
\eeqn
Integrating out $\hat F_9^{\{3,6\}}$ leads to the same result as in (\ref{kinc0}) and
the only difference, as compared to the case without vectors,
appears in the structure of the Chern-Simons terms.
To replace $\hat F_7^{\{2,5\}}$ we add
\beqn \label{lagrm5}
\delta {\cal S} & = & \frac{1}{2\kappa^2_{10}}
\int C_2^{\{1,1\}} \wedge d^{\{1,0\}} \Big( \hat F_7^{\{2,5\}} + B_2^{\{1,1\}} \wedge dC_4^{\{1,4\}} \\
&& \mbox{} \hspace{1cm} + \gamma \Big[ - (\star \omega_3)^{\{2,5\}} - B_2^{\{1,1\}} \wedge (\star \omega_3)^{\{1,4\}}
+ \frac12 (B_2^{\{1,1\}})^2 \wedge (\star \omega_3)^{\{0,3\}} \Big] \Big)\ . \nonumber
\eeqn
Again, integrating out $\hat F_7^{\{2,5\}}$ leads to the old result (\ref{kinc2}).
Further, apart from the kinetic and potential terms given in
(\ref{kinc0}), (\ref{kinc2}) and the second line of (\ref{tdkin2}), we obtain
the following Chern-Simons terms
\beqn
\label{dcs}
{\cal S}_{\rm CS} & = &
\frac{1}{2 \kappa^2_{10}} \int \gamma \Bigg(
\Big( (dC_2)^{\{2,1\}} - C_0 (dB_2)^{\{2,1\}} \Big) \non
&& \quad\quad\quad\quad\quad
\wedge~ \Big( (\star \omega_3)^{\{2,5\}} + B_2^{\{1,1\}} \wedge (\star \omega_3)^{\{1,4\}}
- \frac12 \Big( B_2^{\{1,1\}} \Big)^2 \wedge (\star \omega_3)^{\{0,3\}} \Big) \non
&&
\quad\quad\quad  +~
(dC_0)^{\{1,0\}} \wedge \Big( (\star \omega_3)^{\{3,6\}}
- B_2^{\{1,1\}} \wedge (\star \omega_3)^{\{2,5\}} \non
&&
\quad\quad\quad\quad\quad
-~ \frac12 \Big( B_2^{\{1,1\}} \Big )^2 \wedge (\star \omega_3)^{\{1,4\}}
+ \frac{1}{3!} \Big( B_2^{\{1,1\}} \Big)^3 \wedge (\star \omega_3)^{\{0,3\}} \Big) \Bigg)
\non
&&
+~ \frac{1}{2\kappa^2_{10}}
\int (dC_2)^{\{2,1\}} \wedge (dB_2)^{\{2,1\}} \wedge C_4^{\{0,4\}}\ .
\eeqn
Using (\ref{gamma}), the definition
\be
F_{j\mu \nu} = 2\partial_{[\mu} C_{\nu]j}\ ,
\ee
and the expressions
\beqn \label{omegas}
(\omega_3)_{\mu \nu \rho} & = & 6 A^a_{[\mu} \partial_\nu
A^a_{\rho]} + 2 f^{abc} A^a_\mu A^b_\nu A^c_\rho \ , \non
{(\omega_3)_{\mu \nu}}^{i} & = & 2 A^a_{[\mu} \partial_{\nu]} A^{ai}
+ 2 A^{ai} \partial_{[\mu} A^a_{\nu]} + 2 f^{abc} A^a_\mu A^b_\nu
A^{ci} \ , \non
{(\omega_3)_\mu}^{ij} & = & - 2A^{a[i} \partial_\mu A^{a j]}
+ 2 f^{abc} A^a_\mu A^{bi} A^{cj}\ , \non
(\omega_3)^{ijk} & = & 2 f^{abc} A^{ai} A^{bj} A^{ck}\ ,
\eeqn
one finds
\beqn
&& \frac12 {(\omega_3)_{\mu\nu}}^{i} + B_{[\mu |j|} {(\omega_3)_{\nu]}}^{ij}
- \frac12 B_{[\mu |j|} B_{\nu] k} (\omega_3)^{ijk} \non
&& \hspace{5cm}
=
A^{ai} \tilde{\cal F}_{\mu\nu} + A^{ai} A^{aj} \partial_{[\mu} B_{\nu ]j}
- 2 \partial_{[\mu} \left( A^{ai} \tilde A^{a}_{\nu]} \right)
\ , \non
&& (\omega_3)_{\mu\nu\rho} - 3 B_{[\mu |i|}{(\omega_3)_{\nu\rho]}}^i
+ 3 B_{[\mu |i|}B_{\nu |j|} {(\omega_3)_{\rho]}}^{ij}
- B_{[\mu |i|}B_{\nu |j|}B_{\rho] k} {(\omega_3)}^{ijk}
\non
&& \hspace{5cm}
=~ (\tilde \omega_3 )_{\mu\nu\rho} + 6 A^{ai} \tilde A^a_{[\mu} \partial_\nu B_{\rho] i}
\ ,
\eeqn
where we also used the definitions (\ref{def}) and (\ref{tilda}).
Now one
can express the Chern-Simons terms as
\beqn \label{cstot}
{\cal S}_{\rm CS}
& = & -~  \frac{1}{4 \kappa^2_{10}}
\int d^{10}x \sqrt{-g} \gamma \epsilon^{\mu \nu \rho \sigma} \Big(
\frac12 C_0 \tilde{\cal F}_{\mu \nu}^a \tilde{\cal F}_{\rho \sigma}^a \\
&& \mbox{} \hspace{3.8cm} - \left(F_{j\mu \nu} - C_0 H_{j\mu \nu}\right)
\Big( A^{aj} \tilde{\cal F}_{\rho \sigma}^a
+ \frac12 A^{aj} A^{ai} H_{i\rho \sigma} \Big) \Big) \
\non
&&
+~ \frac{1}{2\kappa^2_{10}}
\int (dC_2)^{\{2,1\}} \wedge (dB_2)^{\{2,1\}} \wedge C_4^{\{0,4\}}\ . \nonumber
\eeqn
As will be pointed out in section \ref{compfer}, this expression matches
the results obtained from supergravity (up to some numerical factors). \\

Let us again see which changes occur in the presence of
3-form flux. As in the case without vectors, (\ref{tdkin2}) shows that
the kinetic term for the scalars $C_4^{\{0,4\}}$ changes and additional
terms in the potential appear. The covariant derivative of
the axions $(C_4)_{ijkl}$, which was (\ref{covder}) when setting
all open string fields to zero, now reads
\be \label{covder2}
\partial_\mu (C_4)_{ijkl}
- 2 (B_2)_{\mu [i} ( dC_2
+2 \gamma \star \omega_3 )_{jkl]}
+ 2 (C_2)_{\mu [i} (dB_2)_{jkl]}
+ \gamma (\star \omega_3)_{\mu ijkl}\ .
\ee
Introducing alternative variables as in (\ref{covderbeta}) one
can rewrite this up to an overall factor as
\beqn \label{dbeta}
D_\mu \beta^{ij} & = &
\partial_\mu \beta^{ij}
+ \frac12 \sqrt{G} (\star dC_2)^{ijk} (B_2)_{\mu k}
- \frac12 \sqrt{G} (\star dB_2)^{ijk} (C_2)_{\mu k} \non
&&  +~ \gamma \sqrt{G} \Big( {(\omega_3)_\mu}^{ij} - (\omega_3)^{ijk} B_{\mu k} \Big) \ .
\eeqn
Using (\ref{omegas}), (\ref{def}) and (\ref{tilda}) we see that
\be \label{aDa}
{(\omega_3)_\mu}^{ij} - (\omega_3)^{ijk} B_{\mu k} = - 2 A^{a[i} \tilde D_\mu A^{aj]}\ .
\ee
Thus (\ref{dbeta}) can alternatively be written as
\be \label{dbeta2}
D_\mu \beta^{ij} = \partial_\mu \beta^{ij}
+ \frac12 \sqrt{G} (\star dC_2)^{ijk} (B_2)_{\mu k}
- \frac12 \sqrt{G} (\star dB_2)^{ijk} (C_2)_{\mu k}
- 2 \sqrt{G} \gamma A^{a[i} \tilde D_\mu A^{aj]}\ .
\ee
Now (\ref{dbeta2}) is invariant under the gauge transformation
(\ref{gaugetrafo2}), whereas invariance under (\ref{gaugetrafo1}) requires
$\beta^{ij}$ to transform according to
\be \label{gaugetrafo3}
\beta^{ij} \rightarrow \beta^{ij} - \frac12 \sqrt{G} (\star dC_2)^{ijk} \epsilon_k\ .
\ee
Similarly, under
\be \label{gaugetrafo4}
(C_2)_{\mu k} \rightarrow (C_2)_{\mu k} + \partial_\mu \epsilon_k
\ee
$\beta^{ij}$ has to transform by
\be \label{gaugetrafo5}
\beta^{ij} \rightarrow \beta^{ij} + \frac12 \sqrt{G} (\star dB_2)^{ijk} \epsilon_k\ .
\ee
In the framework of gauged supergravity this implies that
the same translational isometries of the RR scalars are gauged as
in the case without vectors \cite{hepth0206241,D'Auria:2003jk}.
Demanding invariance of the action under gauge-transformations
of the Kaluza-Klein vectors requires that the Chern-Simons term in the
last line of (\ref{cstot}) is replaced by (\ref{csmod}), analogously
to the case without vectors.
Apart from this modification we do not expect any further changes
to (\ref{cstot}) due to the fluxes, because the open string Chern-Simons
terms do not involve the axions $(C_4)_{ijkl}$ and are
already by themselves invariant under gauge-transformations
of the Kaluza-Klein vectors.\footnote{Ultimately, this is justified
by comparison to the gauged supergravity of \cite{D'Auria:2003jk} and, as mentioned already, our results
match the expressions given in \cite{D'Auria:2003jk} without further modifications.}  \\

On the other hand, the 3-form potential does receive additional
contributions as argued above in the
case without vectors. Comparison with
type IIB theory implies that there is an additional potential term of the form $|H_3^{\{0,3\}}|^2$
and that instead of
$(dC_2 + \gamma \star \omega_3)^{\{0,3\}}$ it is the full
$(dD_2 + \gamma \star \omega_3)^{\{0,3\}}$ that should enter into
$|\hat F_3^{\{0,3\}}|^2$.
The two terms can again be combined to give
\be \label{fullpot}
|(F_3 + \gamma \star \omega_3 )^{\{0,3\}}|^2
+ e^{-2\Phi} |H_3^{\{0,3\}}|^2  =
|(G_3 + \gamma \star \omega_3 )^{\{0,3\}}|^2
\equiv |\hat G_3^{\{0,3\}}|^2 \ .
\ee
Actually, there is a loop-hole in our strategy to combine the
T-dual data and the truncated type IIB action in this case. The
expression (\ref{fullpot}) contains the cross-term $2\gamma C_0
(dB_2)_{ijk} (\star \omega_3)^{ijk}$. This term contains NSNS
3-form flux and open-string fields at the same time and thus
vanishes in both limits. Furthermore, its presence or absence is
not restricted by any symmetry argument. This is different from
the corresponding cross-term in the kinetic term for $\beta^{ij}$,
which is required by KK gauge invariance, cf.\ (\ref{dbeta2}).
Also the shift-symmetry of $C_0$ is broken by the presence of the
D3-branes and can not help to fix the coefficient of $\gamma C_0
(dB_2)_{ijk} (\star \omega_3)^{ijk}$. Nevertheless, we believe that
(\ref{fullpot}) is the correct combination, because it leads to
the same potential as found in the supergravity approach
\cite{hepth0211027,D'Auria:2003jk}, as we will show in a moment. \\

Putting all the pieces together, the bosonic action derived by
T-duality of the type I string contains four different parts
\be \label{totaction}
{\cal S} = {\cal S}_{\rm EH} + {\cal S}_{\rm kin} + {\cal S}_{\rm pot} +
           {\cal S}_{\rm CS}\ .
\ee
In the string frame, they are given in turn by
\beqn \label{skin}
{\cal S}_{\rm EH} & = & \frac{1}{2 \kappa_{10}^2} \int d^{10} x\, \sqrt{-g}
  e^{-2 \Phi} R \ ,
\non
{\cal S}_{\rm kin} & = &
\frac{1}{2 \kappa_{10}^2} \int d^{10} x\, \sqrt{-g}
  e^{-2 \Phi} \left(
  4 \partial_\mu \Phi \partial^\mu \Phi
  - \frac14 G^{ij} H_{i \mu \nu} H_j^{\mu \nu} \right) \non
&& - \frac{1}{4\kappa^2_{10}} \int d^{10}x \sqrt{-g} \left(
  |F_1^{\{1,0\}}|^2 + |F_3^{\{2,1\}}|^2 + |\hat F_5^{\{1,4\}}|^2 \right)
\non
&& - \frac{1}{2\kappa^2_{10}}
  \int d^{10} x\, \sqrt{-g} \gamma e^{-\Phi} \Big(
   G_{ij} g^{\mu \nu} \tilde D_\mu A^{ai} \tilde D_\nu A^{aj} \non
&& \hspace{1.5cm}
+ \frac12 g^{\mu \nu} g^{\rho \sigma} (\tilde {\cal F}^a_{\mu \rho}
                + H_{i\mu \rho}A^{ai})
(\tilde {\cal F}^a_{\nu \sigma} + H_{j \nu \sigma}A^{aj}) \Big)\ , \\
\label{spot}
{\cal S}_{\rm pot} & = &  -\frac{1}{4\kappa^2_{10}} \int d^{10}x
                \sqrt{-g} \Big( |\hat G_3^{\{0,3\}}|^2
+2 \gamma e^{-\Phi} {\rm tr}\, |{\cal F}^{\{0,2\}}|^2 \Big) \ , \\
\label{scs}
{\cal S}_{\rm CS} & = & -  \frac{1}{4 \kappa^2_{10}}
\int d^{10}x \sqrt{-g} \gamma \epsilon^{\mu \nu \rho \sigma} \Big[
\frac12 C_0 \tilde{\cal F}_{\mu \nu}^a \tilde{\cal F}_{\rho \sigma}^a \non
&& \mbox{} \hspace{3cm} - \left(F_{j\mu \nu} - C_0 H_{j\mu \nu}\right)
\Big( A^{aj} \tilde{\cal F}_{\rho \sigma}^a
+ \frac12 A^{aj} A^{ai} H_{i\rho \sigma} \Big) \Big] \non
&&
+~ \frac{1}{2\kappa^2_{10}}
\int (dC_2)^{\{2,1\}} \wedge (dB)^{\{2,1\}} \wedge C_4^{\{0,4\}} \non
&& -~ \frac{1}{4\kappa^2_{10}}  \int C_2^{\{1,1\}} \wedge (dB_2)^{\{2,1\}}
\wedge B_2^{\{1,1\}} \wedge (dC_2)^{\{0,3\}} \non
&& -~ \frac{1}{4\kappa^2_{10}} \int B_2^{\{1,1\}} \wedge (dC_2)^{\{2,1\}}
\wedge C_2^{\{1,1\}} \wedge (dB_2)^{\{0,3\}}\ .
\eeqn


\subsection{Additional modifications from the non-abelian DBI}
\label{secaddmod}

To identify candidate couplings in the effective action that
involve 3-form flux and open string fields at the same time,
one can study the non-abelian D-brane world volume action in
the formulation given in \cite{hepth9910053}. It turns out that 
only the DBI part contains a term that has to be addded to the
effective action derived via T-duality. The CS part in principle has 
the potential to modify the tadpole condition, but we will argue that 
actually it does not. \\

From the expansion of the DBI action for a D3-brane in the
presence of NSNS 3-form flux one deduces a correction to the effective action 
\cite{hepth9910053} that in our conventions reads
\be \label{myersagain}
\frac{i}{3 g_{10}^2 \sqrt{G}} e^{-\Phi} {\rm tr} (A^i A^j A^k) H_{ijk}
~=~ - \frac{1}{2 \kappa^2_{10}} \frac{1}{3!} \gamma e^{-\Phi}
\omega_3^{ijk} H_{ijk}
\ .
\ee
This term represents an extra contribution to the naive abelian
term for the tension of a D3-brane in the presence of NSNS 3-form 
flux.
\\

By analyzing the non-abelian CS action of the D3-branes
one may
have expected a modification of the relevant component of the
equation of motion for $C_4$, since one finds a further coupling
linear in $C_4^{\{4,0\}}$, i.e.\
\be \label{myersfp}
\int {\rm tr} \Big( C_4^{\{4,0\}} \ {\rm i}_{A}{\rm i}_{A} \, B_2 \Big)
~\sim~ \int d^4x\, \sqrt{-g_4} \epsilon^{\mu \nu \rho \sigma}
(C_4)_{\mu \nu \rho \sigma} H_{ijk} \, {\rm tr} (A^i A^j A^k)\ ,
\ee
using the symbolic notation of \cite{hepth9910053}, i.e.\ 
${\rm i}_{A}{\rm i}_{A} \, B_2 = A^j A^i B_{ij}$. This term could 
potentially modify the tadpole cancellation
condition for D3-brane charge. However, one has to take into account 
that the CS action involves RR-fields of 
all degrees and (\ref{myersfp}) is accompanied by another
contribution, the direct analogue of the
term that lead to dielectric D0-branes in \cite{hepth9910053},
\be \label{myersforp}
\int {\rm tr} \Big( {\rm P} \left[ {\rm i}_{A}{\rm i}_{A} \, C_6 \right] \Big)
~\sim~
\int d^4x\, \sqrt{-g_4} \epsilon^{\mu \nu \rho \sigma}
(dC_6)_{\mu \nu \rho \sigma ijk} {\rm tr} (A^i A^j A^k)\ ,
\ee
where ${\rm P}$ denotes the (non-abelian) pull-back.  
The component of $dC_6$ that occurs here is related to the 3-form flux via
duality. To make this more precise, one has to use the
democratic version of the ten-dimensional type I action (already mentioned 
in footnote \ref{demoaction}),
that involves $C_6$ on the same footing as $C_2$ even before
applying the T-duality, cf.\ appendix
\ref{sdactionsec}. This formulation also involves a kinetic
term for the field strength $F_7^{\{4,3\}}$ 
after T-duality. In general, $F_7$ is given by
\be \label{f7}
F_7 = dC_6 + H_3 \wedge (C_4 - \frac12 C_2 \wedge B_2)\ ,
\ee
where we made use of the general formula (see e.g.\ \cite{BKORVP}\footnote{Note that our NSNS $B$-field differs by a
sign from the one used there.})
\be \label{generalf}
\sum_{q=1}^5 F_{2q-1} = \sum_{q=1}^5 d \tilde C_{2q-2} + H_3 \wedge \sum_{q=2}^5 \tilde C_{2q-4}\ ,
\ee
and took into account that our $C_4$ differs from $\tilde C_4$
used in \cite{BKORVP} by $\tilde C_4 = C_4 - \frac12 C_2 \wedge
B_2$, whereas all other $C_p$ coincide, i.e.\ $\tilde C_p = C_p$
for $p \neq 4$. In (\ref{myersfp}) the difference between $\tilde
C_4$ and $C_4$ does not matter because $(B_2)_{\mu \nu}$ and
$(C_2)_{\mu \nu}$ are projected out and their backgrounds vanish
in the vacuum. Now we notice that the two terms (\ref{myersfp})
and (\ref{myersforp}) nicely combine into
\be \label{myersforpres}
\int d^4x\, \sqrt{-g_4} \epsilon^{\mu \nu \rho \sigma}
(F_7)_{\mu \nu \rho \sigma ijk} {\rm tr} (A^i A^j A^k) + \ \cdots\
,
\ee
where the dots stand for further terms, involving also
$C_4^{\{2,2\}}$,
which is dualized in favor of $C_4^{\{0,4\}}$ in the end. We do
not want to go into further details here, but just observe that upon dualizing
$F_7^{\{4,3\}}$ in a similar vein as done in appendix
\ref{sdactionsec}, the term (\ref{myersfp}) disappears from the action and there
is no additional contribution to the tadpole condition for
$C_4^{\{4,0\}}$ after eliminating the superfluous degrees of
freedom. This means that in the tadpole condition (\ref{modtad}) the contribution of
the fluxes to the effective 3-brane charge is
\be \label{nflux}
N_{\rm flux} = \frac{1}{2\kappa_{10}^2 \mu_3}
\int_{\mathbb{T}^6} F_3^{\{0,3\}} \wedge  H_3^{\{0,3\}} \ ,
\ee
exactly as in the case without the open string fields, because the
CS correction from the non-abelian D-brane action drops out. Here
$\mu_3$ denotes the 3-brane string frame tension which for general
$p$-branes is given by\footnote{Note that the tension includes a
factor $\frac{1}{\sqrt{2}}$ as compared to the corresponding
formula of type IIB, as is required in type I and its T-duals, see
\cite{CB} for instance.}
\be \label{itension}
\mu_p = \frac{1}{\sqrt{2}}2\pi (4\pi^2 \alpha')^{-(p+1)/2}\ .
\ee
Moreover, notice that in the self-dual action the kinetic
term for $C_2$ appears with a factor $1/2$ as compared to the
usual type I action. Thus also e.g.\ the 3-form potential after
T-duality would first come with the same additional factor. It is
only after dualizing $F_7^{\{4,3\}}$ that the full 3-form potential, 
as given in (\ref{spot}), is obtained. \\

The fact that the tadpole condition is not changed as compared to
the case without open string fields is important for the
form of the effective four-dimensional potential. First note that
this tadpole constraint descends from the equation of motion of
$(C_4)_{\mu\nu\rho\sigma}$ in the ten-dimensional theory and does
not arise as a dynamical equation in four dimensions. Instead it
needs to be imposed as a constraint that determines the number of
D3-branes. The dual theory constructed in the previous section is
only consistent in the presence of 3-form fluxes if the fluxes do
not contribute any 3-brane charges, i.e.\ one necessarily has
$N_{\rm flux}=0$. This is because T-dualizing pure type I
naturally leads to a theory with 16 D3-branes\footnote{There is a
well-known ambiguity of how to count the individual branes. In a
microscopic description of the T-dual model, there are 64
O3-planes located at the fixed points of $\Theta$, such that local
charge cancellation as in type I is only achieved in the effective
action if all sources are completely smeared out. The
number 16 simply refers to the rank of the gauge group.} that
already fully cancel the charge (and tension) of the O3-planes. A
non-vanishing $N_{\rm flux}$ would then lead to a surplus of
3-brane charge. This will be discussed in more detail in section
\ref{genactsec}, but let us go slightly ahead of things and already mention that
in general the potential receives a further extra contribution
from the tension of the O3-planes and D3-branes, when $N_{\rm
flux}\not =0$. To accommodate this, the first term of (\ref{spot})
can be rewritten using the splitting of $\hat G_3^{\{0,3\}}$ under
$\star$ into imaginary self-dual and anti-self-dual components,
i.e.\ $\star\hat G_3^{\rm ISD} = i
\hat{G}_3^{\rm ISD},\
\star\hat G_3^{\rm IASD} = - i \hat{G}_3^{\rm IASD}$.\footnote{ Let us remind
the reader that we are using a different definition of the Hodge
star than the one used e.g.\ in
\cite{hepth0009211,hepth0105097,hepth0201028}.} One then verifies,
similarly to \cite{hepth0105097}, that the 3-form flux potential
term combines with the extra non-abelian correction to the brane
tension into
\be \label{splitiasd}
\Big( |\hat G_3^{\{0,3\}}|^2 - \frac{4i}{3} \gamma
    e^{-\Phi} {\rm tr} (A^i A^j A^k) H_{ijk} \Big) d{\rm vol}
    = 2 |\hat G_3^{\rm ISD}|^2 d{\rm vol}
        + 2 e^{-\Phi} F_3^{\{0,3\}} \wedge H_3^{\{0,3\}}
\ee
with $d{\rm vol} = \sqrt{G} d^6x$.
Due to the unmodified tadpole condition we see that the second term on 
the right-hand side of (\ref{splitiasd}) is cancelled by the tension
of the localized objects. \\

Note that the absence of the CS correction to the tadpole condition 
is indispensable for producing a positive definite scalar potential 
as is required by matching the results of gauged supergravity 
\cite{hepth0211027,D'Auria:2003jk}.\footnote{Moreover, this form of the 
potential is necessary to be consistent with the 
constraints that derive from the equations of motion, as discussed
in section \ref{sec5}.} In particular, couplings
that could drive a dielectric Myers' effect, which would lead to
non-commutative brane solutions, now appear as the cross
terms in $|\hat G_3^{\rm ISD}|^2$. At a global minimum of the
potential, when $G_3$ is IASD, they therefore cancel out 
\cite{hepth0201028,hepth0211027}.  \\

Let us make a further comment here. The vector couplings
in the kinetic terms (\ref{skin}) are asymmetric in that Chern-Simons corrections
do not occur in the kinetic terms for $C_0$, $C_2^{\{1,1\}}$ and
$B_2^{\{1,1\}}$. This is due to the fact that we did not include the Chern-Simons term
\be \label{cs9branes}
- \mu_9 \int dC_2 \wedge \omega_7
\ee
as part of
\be
\mu_9 \sum_p \int C_p \wedge {\rm ch}({\cal F})
\ee
for the D9-branes in our starting point type I action (\ref{iaction}).
Under T-duality such a term would be mapped according to
\be \label{opencs}
(dC_2)^{\{q,3-q\}} \wedge (\omega_{7})^{\{4-q,q+3\}} ~\mapsto~
- \frac{(-1)^{q(q-1)/2}}{\sqrt{G}} (dD_{2+2q})^{\{q,3+q\}} \wedge
(\star \omega_{7})^{\{4-q,3-q\}}\ ,
\ee
leading, among other things, to terms involving $(dD_8)^{\{3,6\}}$
and $(dD_6)^{\{2,5\}}$ together with the appropriate components of
$\star \omega_7$. Clearly this would modify the dualizing process
described above, leading to Chern-Simons corrections involving
$\star \omega_7$ in the kinetic terms for $C_0$, $C_2^{\{1,1\}}$
and $B_2^{\{1,1\}}$ and to new Chern-Simons terms similar in
nature to (\ref{dcs}) but lengthier. In that case the kinetic
terms appearing in (\ref{skin}) involve the field strengths
\beqn \label{rrsymm}
\hat F_5^{\{1,4\}} &=& (dD_4 +
 \gamma \star \omega_3)^{\{1,4\}} -
   B^{\{1,1\}} \wedge (dD_2 +
 \gamma \star \omega_3)^{\{0,3\}} \ , \non
\hat F_3^{\{2,1\}} &=& (dD_2 - 2 \gamma \star \omega_7)^{\{2,1\}}
  - B^{\{1,1\}} \wedge (dD_0 +
 2 \gamma \star \omega_7)^{\{1,0\}} \ , \non
\hat F_1^{\{1,0\}} &=& (dD_0 +
 2 \gamma \star \omega_7)^{\{1,0\}} \ ,
\eeqn
instead of $F_3^{\{2,1\}}$ and $F_1^{\{1,0\}}$. In this symmetrized version, the coupling
of the truncated type IIB action to the gauge fields, at least for the kinetic
terms, can be summarized by
adding to the $dD_p$ the appropriate component of $\gamma \star \omega_3$ or
$2\gamma \star \omega_7$ depending on the bi-degree of the form.
However, as (\ref{cs9branes}) is a higher order correction in the open string fields and
we restrict ourselves to the lowest order of the DBI action,
we preferred not to include the corrections due to $\omega_7$ in
(\ref{totaction}). \\


\subsection{The action in the Einstein frame}

Finally, in order to
make contact to the supergravity literature, we have to transform
(\ref{totaction}) into the Ein\-stein frame. Transforming to the ten-dimensional
Einstein-frame by rescaling the metric with the string coupling,
$g_{IJ} \rightarrow e^{\Phi/2} g_{IJ}$, leads to
\beqn \label{eini}
2 \kappa_{10}^2 {\cal S} & = &
\int d^{10} x\, \sqrt{-g} \Big[ R - \frac12
  \partial_\mu \Phi \partial^\mu \Phi -\frac14 e^{-\Phi} G^{ij} H_{i \mu \nu} H_j^{\mu \nu} \non
&& \mbox{} \quad\quad
- \frac12 e^{2 \Phi} \partial_\mu C_0 \partial^\mu C_0 - \frac14 e^{\Phi} G^{ij}
  (F_{i\mu \nu} + C_0 H_{i\mu \nu}) (F_j^{\mu \nu} + C_0 H_j^{\mu \nu}) \non
&& \mbox{} \quad\quad
- \frac14  G^{-1} G_{ik} G_{jl} D_\mu \beta^{ij} D^\mu \beta^{kl}
  - \gamma G_{ij} \tilde D_\mu A^{ai} \tilde D^\mu A^{aj} \non
&& \quad\quad
\mbox{} - \frac12 \gamma e^{-\Phi} (\tilde {\cal F}^a_{\mu \nu} + H_{i\mu \nu}A^{ai})
  (\tilde {\cal F}^{a \mu \nu} + H_j^{\mu \nu} A^{aj}) \non
&&  \quad\quad
\mbox{} - \frac12 e^{\Phi} \Big( |\hat G_3^{\{0,3\}}|^2 - \frac{4i}{3} \gamma
    e^{-\Phi} {\rm tr} (A^i A^j A^k) H_{ijk} \Big) \non
&&  \quad\quad
\mbox{} - \frac12 \gamma e^{\Phi}
  G_{ij} G_{kl} f^{abc} f^{ade} A^{bi} A^{ck} A^{dj} A^{el} \Big] \non
&&
\mbox{} -\frac14 \int d^{10} x\, \sqrt{-g_4} \epsilon^{\mu \nu \rho \sigma}
  \Big[ \gamma \sqrt{G} \Big( C_0 \tilde{\cal F}_{\mu \nu}^a \tilde{\cal F}_{\rho \sigma}^a \non
&& \quad\quad
\mbox{} - 2 \left(F_{j\mu \nu} - C_0 H_{j\mu \nu}\right)
  (A^{aj} \tilde{\cal F}_{\rho \sigma}^a
  + \frac12 A^{aj} A^{ai} H_{i\rho \sigma}) \Big) - \beta^{ij} F_{i\mu\nu} H_{j\rho\sigma} \non
&& \quad\quad
\mbox{} - C_{i\mu}B_{j\nu}H_{k\rho \sigma} \sqrt{G} (\star dC_2)^{ijk}
  - B_{i\mu}C_{j\nu}F_{k\rho \sigma} \sqrt{G} (\star dB_2)^{ijk} \Big]\ ,
\eeqn
where we introduced the variables $\beta^{ij}$ and used (\ref{dbeta2}).
This form of the action will be relevant in the section \ref{sec5} when we allow the
bulk fields to vary over the whole internal space, and the open-string fields over the world-volume of the
D-branes, looking for solutions to the ten-dimensional equations of motion. \\

Dimensional reduction of (\ref{eini}) and a Weyl rescaling $g_{\mu\nu} \rightarrow G^{-1/2}g_{\mu\nu}$
to go to the four-dimensional Einstein-frame produces the following effective
action\footnote{We take the background volume to be $(2 \pi \alpha'^{1/2})^6$ and therefore
have $\kappa_4^2 = \kappa_{10}^2 (4 \pi^2 \alpha')^{-3}$. \label{kappa4}}
\beqn \label{effact}
2 \kappa_{4}^2 {\cal S} & = & \int d^4 x\, \sqrt{-g_4} \Big[ R -\frac14 \tg^{ik} \tg^{jl}
\partial_\mu \tg_{ij} \partial^\mu \tg_{kl} - \frac12
\partial_\mu \Phi \partial^\mu \Phi -\frac14 e^{-\Phi} \tg^{ij} H_{i \mu \nu} H_j^{\mu \nu} \non
&& \mbox{} \quad\quad - \frac12 e^{2 \Phi} \partial_\mu C_0 \partial^\mu C_0 - \frac14 e^{\Phi} \tg^{ij}
(F_{i\mu \nu} + C_0 H_{i\mu \nu}) (F_j^{\mu \nu} + C_0 H_j^{\mu \nu}) \non
&& \mbox{} \quad\quad  -\frac14  \tilde G_{ik} \tilde G_{jl} D_\mu \beta^{ij} D^\mu \beta^{kl}
- \tga \tg_{ij} \tilde D_\mu A^{ai} \tilde D^\mu A^{aj} \non
&& \mbox{} \quad\quad  - \frac12 \tga e^{-\Phi} (\tilde {\cal F}^a_{\mu \nu} + H_{i\mu \nu}A^{ai})
(\tilde {\cal F}^{a \mu \nu} + H_j^{\mu \nu} A^{aj}) \non
&& \mbox{} \quad\quad  - \sqrt{G}^{-1} e^{\Phi} |\hat G_3^{\rm ISD}|^2 - \frac12 \tga e^{\Phi}
\tg_{ij} \tg_{kl} f^{abc} f^{ade} A^{bi} A^{ck} A^{dj} A^{el} \Big] \non
&& \mbox{} -\frac14 \int d^4 x\, \sqrt{-g_4} \epsilon^{\mu \nu \rho \sigma}
\Big[ \tga \Big( C_0 \tilde{\cal F}_{\mu \nu}^a \tilde{\cal F}_{\rho \sigma}^a \non
&& \mbox{} \quad\quad  - 2 \left(F_{j\mu \nu} - C_0 H_{j\mu \nu}\right)
(A^{aj} \tilde{\cal F}_{\rho \sigma}^a
+ \frac12 A^{aj} A^{ai} H_{i\rho \sigma}) \Big) - \beta^{ij} F_{i\mu\nu} H_{j\rho\sigma} \non
&& \mbox{} \quad\quad  - C_{i\mu}B_{j\nu}H_{k\rho \sigma} \sqrt{G} (\star dC_2)^{ijk}
- B_{i\mu}C_{j\nu}F_{k\rho \sigma} \sqrt{G} (\star dB_2)^{ijk} \Big]\ .
\eeqn
Here we have already taken into account the tension of the localized objects to get 
the form of the potential (fifth line) and introduced the following notation
\beqn \label{nochmehrtildes}
\tg_{ij} ~=~ \frac{1}{\sqrt{G}} G_{ij}  \ , \quad
\tga ~=~  \gamma \sqrt{G} ~=~ \frac{\kappa_{10}^2}{g^2_{10}}\ .
\eeqn
The action is now in the form that should allow for a direct comparison
to the gauged supergravity theory that captures the
effective dynamics of type I strings with background 3-form fluxes.


\subsection{Comparison to gauged supergravity}
\label{compfer}

In this section, we would like to make explicit the comparison of our
results in (\ref{effact}) to those found in the gauged
supergravity approach, by explaining the translation of
notation, parameters and fields. We view this as strong
independent confirmation that the effective theory, obtained by
modifying the T-dual action in the manner described above to
capture the effects of 3-form fluxes in type I, is a sensible
approximation of the string dynamics in the supergravity regime.
To what extent it is approximate will be subject of section \ref{sec5}. \\

The expressions we are going to compare are the covariant
derivative of the axionic scalars, already discussed in
(\ref{dbeta2}), and the gauge kinetic and Chern-Simons part
(\ref{scs}) of the action (\ref{effact}). Finding basic agreement
(up to two factors of $2$ and a sign, see below) with the
supergravity results of  \cite{D'Auria:2003jk}, see also
\cite{hepth0206241, hepth0207135,hepth0211027} for partial
results, we conclude that the effective models are identical. A
third object of great interest is, of course, the scalar
potential, to which the entire section \ref{sec4} will be devoted.
The expression given in equations (4.97), (4.98) and (5.132) of
\cite{D'Auria:2003jk} for the covariant derivative of the axion
is\footnote{We are using gauge group indices $a, b, ...$ instead
of $i, j, ...$ and $A, B, ...$ instead of $I, J, ...$ to avoid
confusing them with our space-time indices.}
\beqn \label{dbeta3}
D_\mu B^{\Lambda\Sigma} = \partial_\mu B^{\Lambda\Sigma} + f^{\Lambda\Sigma\Gamma}_\alpha A_{\mu\Gamma}^\alpha
  - a^{a[\Lambda} \nabla_\mu a^{a\Sigma]} ,
\eeqn
where the $f_\alpha^{\Lambda\Sigma\Gamma}$, $\alpha=1,2$, are the
numerical parameters for the 3-form fluxes, $A_{\mu\Gamma}^\alpha$
the abelian KK vector fields and the $a^{a\Lambda}$ are scalars
charged under the non-abelian gauge group. Finally, $\nabla_\mu$
denotes their gauge covariant derivative. In view of
(\ref{dbeta2}) this suggests the following mapping of fields
\beqn \label{fieldmap}
B^{\Lambda\Sigma} ~\leftrightarrow~ \beta^{ij} \ , \quad
A_{\mu\Lambda}^1 ~\leftrightarrow~ (B_2)_{\mu i} \ , \quad
A_{\mu\Lambda}^2 ~\leftrightarrow~ (C_2)_{\mu i} \ , \quad
a^{a\Lambda} ~\leftrightarrow~ - A^{ai} \ ,
\eeqn
and flux parameters
\be \label{transflux}
f_1^{\Lambda\Sigma\Gamma} ~\leftrightarrow~  \frac12 \sqrt{G} (\star dC_2)^{ijk} \ , \quad
f_2^{\Lambda\Sigma\Gamma} ~\leftrightarrow~  - \frac12 \sqrt{G} (\star dB_2)^{ijk} \ ,
\ee
where we have set $\tilde \gamma = 1$ throughout.
The signs are going to become clearer below.
The non-abelian vectors should map to the $\tilde A^a_\mu$ of (\ref{tilda}).
Using this map leads to agreement between our result (\ref{dbeta2}) and (\ref{dbeta3})
(up to a factor of $2$ for the last term).
\\

From the gauge kinetic and CS part of (\ref{effact}) we can read off the $\theta$-angles and coupling constants of the
different gauge fields, which are already present in the ungauged theory and which we define as the coefficients in front of
${F}_{\mu\nu} {F}^{\mu\nu}$ or $\epsilon^{\mu\nu\rho\sigma} {F}_{\mu\nu} {F}_{\rho\sigma}$, respectively.
$F_{\mu\nu}$ now stands for any
kind of gauge field strength, ${\cal F}^a_{\mu\nu}$, $F_{i\mu\nu}$ or $H_{i\mu\nu}$.
We put them into a matrix labelled by $(a,\alpha,i)$ and read off
\beqn \label{theta}
&&
\theta^{ab} ~ = ~ -\frac12 C_0 \delta^{ab}\ , \quad
\theta^{i1a} ~ = ~ A^{ai}\ , \quad
\theta^{i2a} ~ = ~ - C_0 A^{ai}\ , \quad
\theta^{i1j1} ~ = ~ 0\ , \non
&& \theta^{i2j2} ~ = ~ -\frac12 C_0 A^{ai} A^{aj}\ , \quad
\theta^{i1j2} ~ = ~ \theta^{i2j1} ~ = ~ \frac14 \beta^{ij} + \frac14 A^{ai} A^{aj}\
\eeqn
and
\beqn \label{gaugecouplings}
&&
(g^{-2})^{ab} ~ = ~ - e^{-\Phi} \delta^{ab}\ , \quad
(g^{-2})^{i1a} ~ = ~ 0 \ , \quad
(g^{-2})^{i2a} ~ = ~ - 2 e^{-\Phi} A^{ai} \ ,  \non
&&
(g^{-2})^{i1j1} ~ = ~ - \frac12 e^{\Phi} \tg^{ij}\ ,\quad
(g^{-2})^{i2j2} ~ = ~ - \frac12 (e^{-\Phi} + e^{\Phi} C_0^2) \tg^{ij} - A^{ai} A^{aj} e^{-\Phi} \ , \non
&&
(g^{-2})^{i1j2} ~ = ~ (g^{-2})^{i2j1} =  - \frac12 C_0 e^{\Phi} \tg^{ij}\ ,
\eeqn
leaving a factor $1/(4\kappa_4^2)$ in front of the action.
The corresponding supergravity results can be found in (5.130) together
with (3.90)\footnote{Beware a misprint in an older version,
where in the first line of (3.91) the last term
should contain a factor of $i$ instead of $\frac{i}{2}$.}
of \cite{D'Auria:2003jk} in the form of a matrix ${\cal N}$ of complex
coupling constants. Using
\be \label{fplusminus}
{F}^\pm_{\mu \nu} = \frac12 ({F}_{\mu \nu} \pm \frac{i}{2} \epsilon_{\rho\sigma\mu\nu} {F}^{\rho \sigma}) \ ,
\ee
%
and suppressing indices one has
\beqn \label{calns}
-i ({\cal N} {F}^+_{\mu \nu} {F}^{+ \mu \nu}
- \bar{\cal N} {F}^-_{\mu \nu} {F}^{- \mu \nu})
& = & {\rm Im}({\cal N}) {F}_{\mu \nu} {F}^{\mu \nu}
+\frac12 {\rm Re}({\cal N}) {F}_{\mu \nu} {F}_{\rho \sigma}
\epsilon^{\mu \nu \rho \sigma}\ . \nonumber
\eeqn
Thus the $\theta$-angles and couplings are given by\footnote{Note the extra factor of $2$
in the second line of (5.130) in \cite{D'Auria:2003jk}.}
\beqn
&&
{\cal N}^{ab} ~\leftrightarrow~  2 \theta^{ab} + i (g^{-2})^{ab} \ , \quad
{\cal N}^{i\alpha a} ~\leftrightarrow~  \theta^{i\alpha a} + \frac{i}{2} (g^{-2})^{i\alpha a} \ , \non
&&
{\cal N}^{i\alpha j\beta} ~\leftrightarrow~  2 \theta^{i\alpha j\beta} + i (g^{-2})^{i\alpha j\beta} \ .
\eeqn
The expressions one finds for the entries of ${\cal N}$
are identical to those in (\ref{theta}) and (\ref{gaugecouplings}) upon identifying
\beqn \label{dict}
&&
C ~ \leftrightarrow ~ C_0\ , \quad
\varphi ~ \leftrightarrow ~ -\Phi\ , \quad
g^{\Lambda\Sigma} ~ \leftrightarrow ~ \frac12 \tg^{ij}\ ,
\eeqn
in addition to (\ref{fieldmap}), up to the overall factor of $1/(4\kappa_4^2)$, the
sign in the last line
of (\ref{gaugecouplings}) and a factor of $2$ in the $\beta^{ij}$-dependent term in
(\ref{theta}). To complete
the translation of fields, the $L_\alpha$ in the $SU(1,1)/U(1)$
coset are translated into our $SL(2,\mathbb{R})/U(1)$ scalars
$C_0,
\Phi$ by
\beqn \label{cosetscal}
L^1
~ \leftrightarrow ~ - \frac{i}{\sqrt{2}} e^{\Phi/2} \ , \quad
L^2
~ \leftrightarrow ~ \frac{i}{\sqrt{2}} e^{\Phi/2} \left( C_0 + i e^{-\Phi} \right)
\eeqn
with $L_1 = L^2,\ L_2 = - L^1$.


\section{The potential}
\label{sec4}

From the fifth line of (\ref{effact}) we can read off the effective four-dimensional
potential
\beqn \label{effpot}
{\cal V}_{\rm eff} &=&
 \frac{ e^\Phi}{2\kappa^2_{4} \sqrt{G}} |(G_3 + \gamma\star\omega_3)^{\rm ISD}|^2
 + \frac{\tilde\gamma e^\Phi}{4\kappa_{4}^2}
\tg_{ij} \tg_{kl} f^{abc} f^{ade} A^{bi} A^{ck} A^{dj} A^{el}\ .
\eeqn
Before discussing the implications of this potential let us compare it to the expression
derived with the formalism of ${\cal N} =4$ gauged supergravity.
In the conventions of \cite{D'Auria:2003jk}
the scalar potential was written\footnote{We absorb factors of $p!$ into our
definition of the norm (\ref{defforms}).}
\be \label{ferrarai}
{\cal V}_{\rm SUGRA} = \frac{1}{2} | F^{ABC -} + C^{ABC -}|^2
 + \frac14 | L_2 f^{abc} q^{bA} q^{cB} |^2 \ ,
\ee
where $F$ is the 3-form flux
\be
F^{ABC} = L^\alpha f^{ABC}_\alpha = L^\alpha f^{\Lambda\Sigma\Gamma}_\alpha E^A_\Lambda E^B_\Sigma E^C_\Gamma\ ,
\ee
the upper index $F^-$ referring to the ISD part, $C^{ABC}$ is the CS correction
\be
C^{ABC} = L_2 f^{abc} q^{aA} q^{bB} q^{cC}
= L_2 f^{abc} a^{a\Lambda} a^{b\Sigma} a^{c\Gamma} E^A_\Lambda E^B_\Sigma E^C_\Gamma\ ,
\ee
and indices are pulled back with the vielbein $E^A_\Lambda$ which is related to the metric
(\ref{nochmehrtildes}) via (\ref{dict}) and
\be
g_{\Lambda\Sigma} = \delta_{AB} E^A_\Lambda E^B_\Sigma \ .
\ee
The metric moduli together with the axions from $B^{\Lambda\Sigma}$ and the
$a^{a\Lambda}$ parametrize the scalar manifold
\be
\frac{SO(6,6+N)}{SO(6) \times SO(6+N)} \ .
\ee
The identification of the two expressions is then accomplished by applying
(\ref{fieldmap}), (\ref{transflux}) and (\ref{dict}), where one has to take care that the extra signs from
$\star\star =-1$ and (\ref{cosetscal}) cancel out. We find agreement ${\cal V}_{\rm SUGRA} = {\cal V}_{\rm eff}$
after setting $\kappa^2_4 = \tilde\gamma = 1$ in (\ref{effpot}).
Strictly speaking, this choice of parameters
is allowed only within supergravity. String theory relates the two in a way that is not
consistent with setting them equal.\\

An important phenomenological feature of (\ref{effpot}) is that it involves
the dilaton not just as a prefactor, so one can hope to stabilize the string
coupling. Concentrating on Minkowski vacua we require
\be
|(G_3 + \gamma\star\omega_3)^{\rm ISD}|^2 = 0 \ , \quad
{\rm tr}|{\cal F}^{\{0,2\}}|^2 = 0 \ .
\ee
As ${\rm tr}|{\cal F}^{\{0,2\}}|^2 = 0$ also implies vanishing of $\omega_3$,
the relevant term for fixing $\Phi$ is given by
$|G_3^{\rm ISD}|^2 = 0$, \cite{hepth0211027}. The condition for $G_3$ to be IASD can be written
\be \label{dilfix}
\star F_3 - e^{-\Phi} H_3 = 0
\ .
\ee
Rewriting this via (\ref{transflux}) one finds a relation
\be
f_1^{\Lambda\Sigma\Gamma} - C_0 f_2^{\Lambda\Sigma\Gamma} + e^{-\Phi} (\star f_2)^{\Lambda\Sigma\Gamma} = 0 \ ,
\ee
where $\star f_2$ is defined with respect to the metric $g_{\Lambda\Sigma}$.
Setting $C_0=0$ for simplicity, we see that
%
\be \label{dilfixcond}
f_1^{\Lambda\Sigma\Gamma} + e^{-\Phi} \frac{1}{3!} \sqrt{{\rm det}(g_{\Lambda\Sigma})}\,
  g^{\Lambda\Lambda'} g^{\Sigma\Sigma'} g^{\Gamma\Gamma'}
  \hat\epsilon_{\Lambda'\Sigma'\Gamma'\Delta\Pi\Omega} f_2^{\Delta\Pi\Omega} =0  \ ,
\ee
where $\hat\epsilon$ is just again the antisymmetric symbol with value $\pm 1$.
It appears that in general the dilaton becomes a function of the
metric moduli that also depends on the choice of fluxes.
It was argued in \cite{D'Auria:2003jk} that the 3-form fluxes allowed by the
gauging, i.e.\ consistent with a supersymmetric Lagrangian,
can be put into a form such that, in complex coordinates,
all components $f_\alpha^{m\bar mn}$ and $f_\alpha^{m\bar m\bar n}$ vanish.
We will see in section \ref{trunc} that this ensures a fixing of the
dilaton at least in vacua in which supersymmetry has been broken to ${\cal N}=1$ via a super-Higgs-mechanism.


\subsection{The role of the superpotential}

In compactifications of type IIB theory
on Calabi-Yau spaces with fluxes, the scalar potential
that descends from the kinetic term of $G_3$, after breaking supersymmetry to ${\cal N}=1$ either
by orientifolding \cite{hepth0105097} or by taking a certain decompactification limit as in
\cite{Michelson:1996pn,Taylor:1999ii},
can be expressed in terms of the superpotential
\be \label{supoti}
W_{\rm flux} = \int_{{\cal M}_6} G_3 \wedge \Omega_3\ ,
\ee
proposed by independent arguments in \cite{Gukov:1999ya,Gukov:1999gr},
with $\Omega_3$ the holomorphic 3-form. In that case,
the scalar potential only depends on the complex structure moduli of the Calabi-Yau
and the dilaton. As is well known, in the heterotic or type I string the superpotential
gets a further contribution involving the scalars descending from the ten-dimensional
vectors \cite{ED1,Ferrara:1986qn,ED}
\be \label{supotheti}
W_{\rm het/I} = \int_{{\cal M}_6} \omega_3 \wedge \Omega_3\ .
\ee
As in the ${\cal N}=1$ case, also for ${\cal N}=4$ the potential
descends from the ten-dimensional kinetic terms and
can thus be written in the form of $|\hat G_3^{\rm ISD}|^2$ plus tr$|{\cal F}^{\{0,2\}}|^2$,
up to constants. Therefore, one might wonder whether it is still  possible to express the potential
in terms of a ``superpotential''.
In view of the effective scalar potential (\ref{effpot})
we have found in type I$'$ with extra 3-form flux, it seems more appropriate to
consider the ``superpotential''
\be \label{supot}
W_{{\rm I}'} = \int_{{\cal M}_6} (G_3 + \gamma\star\omega_3) \wedge \Omega_3 \ .
\ee
However, in the toroidal compactification the potential also depends on the K\"ahler form, which
reflects the fact that the moduli space of a torus does not split into a direct
product of complex structure and K\"ahler moduli as on a Calabi-Yau. On a
Calabi-Yau manifold, the ISD 3-forms are the $(1,2)$- and $(3,0)$-forms,
see e.g. \cite{hepth0105097}.
Moreover, all $(1,2)$-forms are primitive \cite{WELLS}.
On the torus, on the other hand, there are non-primitive $(1,2)$-forms
of the form $J \wedge d \bar z^m$ where $J$ is the K\"ahler form, and these are also IASD.
Moreover, there are further ISD $(2,1)$-forms of the form
$J \wedge d z^m$ \cite{hepth0201028} which all
enter into the scalar potential (\ref{effpot}). These facts make it very cumbersome to express the potential
in terms of the superpotential (\ref{supot}).\footnote{In order to
follow the calculation in the Calabi-Yau case
\cite{Michelson:1996pn,Taylor:1999ii,hepth0105097,hepth0107264}, one would have to know
what should replace the formula $D_{Z^\alpha} \Omega_3 = \chi_\alpha$, that gives a
basis for the $(2,1)$-forms $\chi_\alpha$ in terms of covariant derivatives
of $\Omega_3$ with respect to the complex structure moduli $Z^\alpha$.
For the torus, one would need a corresponding formula
giving a basis of only the primitive $(2,1)$-forms.
However, the split of the $(2,1)$-forms into primitive and
non-primitive ones depends on the moduli.} \\

Still, the superpotential (\ref{supoti}) has been used in the literature
to encode the conditions for unbroken supersymmetry in a toroidal
(or K3$\times \mathbb{T}^2$) background with 3-form flux
\cite{hepth0201028,Tripathy:2002qw}.\footnote{The case of K3$\times \mathbb{T}^2$
has also been analyzed from the point of view of gauged supergravity in \cite{Andrianopoli:2003jf}.}
There it was shown that demanding supersymmetry is strong enough to
fix the period matrix of the torus completely, which implies fixing its
complex structure. The conditions
for supersymmetry are stronger than demanding extremality of the potential, but
are equivalent to the extremality conditions of the
superpotential (\ref{supoti}).
As only ISD 3-form flux enters the potential, there are possibilities to turn on fluxes
without generating a vacuum energy. The IASD 3-forms consist in
$G_3$ being a $(0,3)$-, or a primitive $(2,1)$-form
or of the type $J \wedge d\bar{z}^m$. On the other hand,
in order to preserve supersymmetry the flux has to be a primitive $(2,1)$-form
\cite{hepth0009211,hepth0010010} for at least one complex structure. The number of
unbroken supersymmetries depends on the number of complex structures for which this
condition is fulfilled \cite{hepth0201028,hepth0201029}. Moreover, it is obvious
from the formulas of the gravitino masses given in \cite{hepth0206241} that turning on
any non-trivial 3-form flux breaks supersymmetry at least to ${\cal N} = 3$.
Since the coupling to the open string
fields is manifest in the supervariation of the gravitinos, dilatinos and
gauginos by replacing $F^{ABC}$ with $F^{ABC}+C^{ABC}$ as used in (\ref{ferrarai})
\cite{D'Auria:2003jk}, it is to be expected
that the supersymmetry conditions in the coupled system are also captured by
the modified superpotential (\ref{supot}) (in addition to the ``D-term'' appearing in the
second contribution to the potential (\ref{ferrarai})), although the scalar potential
(\ref{effpot}) cannot be expressed through $W_{{\rm I}'}$ in an obvious way.


\subsection{Truncating to ${\cal N}=1$}
\label{trunc}

Regarding the discussion of the superpotential (\ref{supot})
in the previous section, it appears interesting to study the breaking
of supersymmetry in the present model from ${\cal N}=4$ to smaller
numbers of supercharges, especially to ${\cal N}=1$, in which case the
potential should be expressible in terms of a superpotential (and a D-term).
This has been done to some extent in the framework
of supergravity \cite{hepth0110277,hepth0202116,hepth0206241,hepth0207135,hepth0211027}
and in this section we would simply like to make contact to \cite{hepth0207135}. \\

By turning on more and more components of the 3-form
flux one can successively break supersymmetry from
${\cal N}=4$ to ${\cal N}=1$,
as has been described in \cite{hepth0206241,hepth0207135}.
In the super-Higgs effect, the decoupling of massive modes is restricted by the
requirement that they must fill massive representations of the surviving supersymmetry
algebra. If one breaks supersymmetry from ${\cal N}=4
\rightarrow 3 \rightarrow 2 \rightarrow 1$ successively, 6, 10 or all 12 (Kaluza-Klein) vector fields
get massive via their St\"uckelberg couplings (\ref{dbeta}).
This is prescisely as required by a successive decoupling of massive $3/2$ multiplets
under ${\cal N}=3,2,1$ supersymmetry, which eat just the 6, 4 and 2
vectors, respectively, those that disappear from the massless sector.
In this successive supersymmetry breaking the matter content of (\ref{spectrum}),
a spin-2 and six spin-1 multiplets of ${\cal N}=4$, leads to a spin-2 and three chiral
multiplets of ${\cal N}=1$ \cite{hepth0207135}.
Therefore, 3 gravitini, all 12 vectors,
18 out of 21 scalars $\tilde G_{ij}$, 12 out of 15 axions $\beta^{ij}$ and $\tau$
get masses with suitable degeneracies
to fill massive ${\cal N}=1$ multiplets. The remaining three complex scalars are
given by the diagonal components $\tilde G_{1\bar 1},\, \tilde G_{2\bar 2},\, \tilde G_{3\bar 3}$
of the now hermitian metric and the appropriate components of $\beta^{ij}$.
Together with the open string
moduli they parametrize the remnant scalar manifold
\be
\frac{SU(1,1)}{U(1)} \times
\frac{SO(6,6+N)}{SO(6)\times SO(6+N)}\quad \rightarrow \quad
\left( \frac{SU(1,1+N)}{U(1)} \right)^3 \ .
\ee
In this particular situation one can convince oneself that the condition
(\ref{dilfix}) is really sufficient to fix the dilaton. The square root of the
determinant of the metric is the product
$\tilde G_{1\bar 1} \tilde G_{2\bar 2} \tilde G_{3\bar 3}$ of the
real parts of the only remaining three moduli, which drop out
from (\ref{dilfixcond}) if the fluxes are restricted as explained below (\ref{dilfixcond}). Thus
(\ref{dilfix}) becomes independent of the metric and $\Phi$ is fixed to some
rational value once the fluxes $F_3$ and $H_3$ are subjected to Dirac
quantization. A natural value for the string coupling
would appear to be of order one, but moderately
small numbers are also easily accessible.
In \cite{hepth0211027} it was suspected that in the process of
integrating out the dilaton, the first term of the potential (\ref{effpot}) vanishes,
such that the potential of the effective ${\cal N}=1$ theory used in \cite{hepth0207135}
is solely given by the second term with a constant value for the dilaton. We were not able to
perform this integrating out explicitly, but assume in the following that the
${\cal N}=1$ potential is only given by the second term of (\ref{effpot}).
\\

It is well known in the literature how to obtain the ${\cal N}=1$
potential from a dimensional reduction of the kinetic term for the
vector fields \cite{ED1,Ferrara:1986qn}.
From (\ref{effpot}) (and setting $2 \pi \alpha'^{1/2}=1$) we extract
\beqn \label{gaugepotential}
4 g_{10}^2 {\cal V}_{\rm {\cal N}=1} = e^{\Phi} \tilde G_{ik} \tilde G_{jl} {\cal F}^{aij}{\cal F}^{akl} =
e^{\Phi} \tilde G_{ik} \tilde G_{jl} f^{abc} f^{ade} A^{bi} A^{cj} A^{dk} A^{el}\ ,
\eeqn
and it is understood that the dilaton and
all metric moduli except the diagonal ones are frozen in the ${\cal N} = 1$ vacuum.
The potential (\ref{gaugepotential}) can be split into an F-term and a D-term by
identifying the metric components with the (real parts of the) K\"ahler coordinates $t_m$ in the following way
\be
t_m + \bar t_{\bar m} - C^{am} \bar C^{a\bar m} = (\tilde G_{m \bar m})^{-1} \ .
\ee
Here we have introduced the complex fields
\be \label{cam}
C^{am} = \frac{1}{\sqrt2} (A^{a (2m-1)} + i A^{a (2m)})\ .
\ee
Using the Bianchi identity (which implies
${\cal F}^a \wedge {\cal F}^a = 0$),
one can rearrange terms to arrive at
\be \label{gaugepotentialsplit}
\tilde G_{ik} \tilde G_{jl} {\cal F}^{aij}{\cal F}^{akl} =
2 \tilde G_{m \bar n} \tilde G_{o \bar p} \left( {\cal F}^{amo} {\cal F}^{a \bar n \bar p} +
                   {\cal F}^{a m \bar n} {\cal F}^{a o \bar p} \right)\ .
\ee
The $(2,0)$ part can be written\footnote{Note the misprint in formula (12) of
\cite{hepth0207135}.}
\be \label{dbifterm}
\tilde G_{m\bar n} \tilde G_{o\bar p} {\cal F}^{amo}{\cal F}^{a\bar n\bar p} =
e^{\cal K} \sum_{m=1}^3 \left( t_m + \bar t_{\bar m} - C^{am} \bar C^{a \bar m} \right)
 \frac{\partial W}{\partial C^{bm}}
 \frac{\partial \bar W}{\partial \bar C^{b \bar m}}\ ,
\ee
with superpotential
\be \label{superpotdbi}
W = \frac{1}{3!} f^{abc} C^{am} C^{bn} C^{co} \hat \epsilon_{mno}\ ,
\ee
where the hat on the epsilon symbol was explained below (\ref{tdrr2}),
and K\"ahler potential
\be
{\cal K} = -\sum_{m=1}^3{\log \left( t_m + \bar t_{\bar m}
                                          - C^{am} \bar C^{a \bar m} \right) }\ .
\ee
That the potential-term (\ref{dbifterm}) is of the usual form
$e^{\cal K}(K^{\alpha \bar \alpha} D_\alpha W \bar D_{\bar \alpha} \bar W - 3 |W|^2)$, where $\alpha$ denotes
all chiral fields, $t^m$ and $C^m$, and $K^{\alpha \bar \alpha}$ the inverse of the
K\"ahler metric, was shown in \cite{hepth0207135}.
The fact that the $-3 |W|^2$ drops out in (\ref{dbifterm}) is due
to the no-scale structure of the potential.
The choice of complex coordinates implies a diagonal period matrix, and the holomorphic
3-form can be written $\Omega_3 = dz^1 \wedge dz^2 \wedge dz^3$, so that
(\ref{superpotdbi}) can be expressed as in (\ref{supotheti}).\footnote{Note that one could have defined
the variables $C^{am}$ of (\ref{cam}) alternatively as $C^{am} = \frac{1}{\sqrt2} (A^{a (2m-1)} - i A^{a (2m)})$.
In that case, the superpotential (\ref{superpotdbi}) would have been of the form (\ref{supot}),
with vanishing $G_3$.}
In a similar way one can identify the
$(1,1)$ component of (\ref{gaugepotentialsplit}) with an ${\cal N}=1$
D-term
\be \label{dterm}
\tilde G_{m\bar n} \tilde G_{o\bar p} {\cal F}^{a m\bar n}{\cal F}^{a o \bar p} =
\left( \sum_{m=1}^3 \frac{1}{t_m + \bar t_{\bar m} - C^{em} \bar C^{e \bar m}}
      f^{bcd} C^{cm} \bar C^{d \bar m}\right)^2 \ .
\ee
Interestingly, adding a world-volume gauge flux by hand, i.e.
adding a constant $f^{am\bar n}$ to ${\cal F}^{am\bar n} = f^{abc}
C^{bm} \bar C^{c \bar m}$,
would lead to a Fayet-Iliopolous term in (\ref{dterm}). \\


\section{Vacua with fluxes and warped metric}
\label{sec5}

In the previous sections we have employed T-duality to transform the effective action
obtained by compactifying type I strings on a torus and neglecting all dependence of
the fields on internal directions. The additional terms in the effective action
with NSNS 3-form flux on the internal space were added in via comparison with type IIB.
However, constant fields will in general no longer
solve the ten-dimensional equations of motion once fluxes are turned on. In this section we shall
look for solutions to these equations, based on the action (\ref{eini}),
with more general background configurations and non-trivial profiles
admitting four-dimensional Minkowski vacua. As explained in the introduction, in order
to derive the equations of motion we allow for a depence of the bulk fields on the
internal coordinates and further modify the action (\ref{eini}) by introducing
a background for the 5-form field strength commensurate with the T-dual
world sheet parity projection (\ref{modwsp}) of type IIB. Moreover, we take into account that
the D3-branes and O3-planes T-dual to the D9-branes and O9-planes are actually
localized objects. We consider the modified equations of motion as describing
the coupling of the bulk fields to the tension of the branes (respectively\ O-planes)
and their world-volume fields. We find that the gauge fields
only cause minor corrections to the known solutions
for a background with vanishing world-volume fields \cite{hepth0009211,hepth0105097}. For these
solutions it has been argued that constant internal fields may still be
considered approximate solutions in the large volume limit where the warp factor may
be taken to be approximately constant, thus a posteriori justifying the
effective action (\ref{effact}) --- at least in this limit. \\

\subsection{Generalized ansatz and modified action}
\label{genactsec}

Our starting point is the ten-dimensional action (\ref{eini}) in which we
now allow the bulk fields to depend on the internal coordinates. Furthermore,
we have to implement some modifications, that we discuss in the following. \\

As was already explained, since the modified world sheet projection
$\Omega\Theta(-1)^{F_L}$ does not project out fields locally, but only demands
symmetry or antisymmetry of the background, we may also allow non-trivial
backgrounds for non-dynamical fields, as long as they respect Poincar\'e invariance.
This refers to the 3-form fluxes $G_3^{\{0,3\}}$, which are not field strengths of
any dynamical potentials but do survive the projection, and also to the
components $F_5^{\{0,5\}}$ and $F_5^{\{4,1\}}$ of the 5-form.
The latter is also subject to the self-duality constraint, such that both
5-form components can be parametrized through a single function,\footnote{Note that due to
our different definition of the Hodge star the internal components have the
opposite sign as in \cite{hepth0009211}.}
\be \label{5-formansatz}
(F_5)_{\mu_1\, ... \, \mu_4 i} =
\frac{1}{\sqrt{g_4}} \epsilon_{\mu_1\, ... \, \mu_4}
  \partial_i \alpha \ , \quad
(F_5)_{i_1 \, ... \, i_5} =
 \frac{1}{\sqrt{g_4}} \epsilon_{i_1\, ... \, i_5 j} (\partial_k \alpha) g^{jk} \ ,
\ee
where $\alpha = \alpha(x^i)$ is antisymmetric under $\Theta$, since
$\Omega(-1)^{F_L}$ leaves $C_4^{\{4,0\}}$ invariant.
In the equations of motion, this background enters in the same way
as in type IIB.
For the metric we use the general warped product ansatz
\be \label{genwarp}
ds_{10}^2 = g_{IJ} dx^I dx^J
 = \Delta(x^k)^{-1} \hat g_{\mu\nu} dx^\mu dx^\nu +
   \Delta(x^k)^b \hat g_{ij} dx^i dx^j \ .
\ee
The factor in front of the internal part is purely conventional,
but if one intends to identify $\hat g_{ij}$ with a Ricci-flat
or even constant metric eventually, it can be helpful to make the
warping explicit. In general, one may also want to consider a non-trivial internal
profile for the axion $F^{\{0,1\}}_1$, as was done in \cite{GP} to
analyze the effects of 7-branes, but we shall not do so here.
Furthermore, in looking for a solution to the coupled system of equations we
restrict our ansatz to the case of vanishing field strengths $F_3^{\{2,1\}}$
and $F^{\{1,0\}}_1$. \\

Second, the open string fields should be allowed to vary only over
the world-volume of the D-branes present in the background. As
long as all fields were constant on the internal space, there was
no distinction between fields localized on some brane and fields
propagating throughout the bulk. T-duality of type I string theory
produced (\ref{tdym}), where the kinetic terms for the gauge
fields only involve the determinant of the four-dimensional
metric, as for gauge fields localized on a D3-brane, but there is
no delta-function to localize the fields. Thus, (\ref{tdym})
refers to ``smeared out'' D3-branes. In the same way, there was no
localization of CS interactions involving open string fields.
Further, once we have introduced localization, the tensions of
D-branes and O-planes do not cancel locally anymore, and we need
to make the tension terms explicit. O3-planes and D3-branes appear
naturally in the toroidal model, since they are just the images of
the O9-planes and D9-branes of type I under T-duality. In fact,
one could also try to introduce higher-dimensional branes, say
D7-branes, and have non-trivial gauge fluxes on their
world-volume, which induces effective D3-brane charges. We refrain
from doing so here, and come back to this option later. One of the
main reasons is that adding world-volume gauge fluxes would
require us to take higher terms in the DBI effective action into
account, which would restrict us to abelian gauge fields, since
the non-abelian DBI action is not known to higher order. \\

From expanding the DBI action to second order in the gauge field strength one
infers the relative factor between the tension term and the terms of the
world-volume action already present in (\ref{eini})
to be $\frac12(2 \pi \alpha')^2$. Setting $2 \pi \alpha'=1$ we thus use for the
D3-brane action\footnote{Note that this is consistent with (\ref{effact}) as
$\frac12(2 \pi \alpha')^2 \mu_3 = \tilde \gamma (2 \kappa_4^2)^{-1}$, where $\kappa_4$
is given in footnote \ref{kappa4} and we used (\ref{itension}) and
$g_{10}^2 = \sqrt{2} (2 \pi)^7 \alpha'^3$, cf. \cite{JUPP}.}
\beqn \label{d3local}
{\cal S}_{\rm D3} &=& - \mu_3 \int d^{10}x\
  \sqrt{-g_4} \Bigg( \sum_{\rm branes} \Big(s_m - \frac{i}{3} {\rm tr} (A^i A^j A^k) H_{ijk}\Big) \delta^{6}( x_m - \bar x_m) \non
&& \quad +~
  \frac{1}{2}
  \sum_{\rm branes}  s_m \delta^{6}( x_m - \bar x_m)
 \Big( g_{ij} g^{\mu \nu} \tilde D_\mu A^{ai} \tilde D_\nu A^{aj} \non
&& \hspace{3cm} +~
 \frac12 e^{-\Phi} g^{\mu \nu} g^{\rho \sigma} (\tilde {\cal F}^a_{\mu \rho} + H_{i\mu \rho}A^{ai})
(\tilde {\cal F}^a_{\nu \sigma} + H_{j \nu \sigma}A^{aj})  \non
&& \hspace{3cm} +~  \frac12 e^{\Phi}
 g_{ij} g_{kl} f^{abc} f^{ade} A^{bi} A^{ck} A^{dj} A^{el} \Big) \Bigg)\ .
\eeqn
The additional term from the non-abelian correction to the DBI
action has also been included above. Just to keep track of the
overall sign we have introduced coefficients $s_m=\pm 1$ for the
tension of the branes. In order to
avoid kinetic terms with the wrong sign for the world volume
fields, one would choose $s_m=1$. In the background, the only non-vanishing terms are the
tension in the first line and the scalar potential tr$|{\cal
F}^{\{0,2\}}|^2$ in the last. The sum over branes now also implies
that the gauge fields are labelled individually for any single
stack of branes, but we do not want to encumber the notation with
another index. Since we have made the D-brane tension explicit we
also have to add the O3-plane tension
\be \label{o3tension}
{\cal S}_{\rm O3} = - \mu_3 Q_3 \int d^{10}x\
  \sqrt{-g_4} \sum_{k=1}^{64} \delta^{6}( x_k - \bar x_k)
\ee
for the 64 planes localized at the fixed points $\bar x_k$ of $\Theta$ and $Q_3 = -1/4$.
Here we assume that the O3-planes have standard negative tension and negative charge.
The total action for the open string sector including the O3-planes is then
given by
\be \label{op2}
{\cal S}_{\rm op} =
{\cal S}_{\rm D3} +
{\cal S}_{\rm O3} \ .
\ee
In the same fashion, the CS interactions with open string fields have to be localized.
For instance, when we write $|\hat F_3|^2$, we now understand
\be \label{cslocal}
|\hat F_3|^2 d{\rm vol} = |dD_2|^2 d{\rm vol} + \sum_{\rm branes}
q_m (2 \pi)^6 \delta^6( x_m -\bar x_m) \Big( 2 \gamma (\star \omega_3) \wedge * dD_2
+ \gamma^2 |\star \omega_3|^2 d{\rm vol} \Big) \ ,
\ee
with $d{\rm vol} = d^{10}x \sqrt{-g}$.
The $q_m = \pm 1$ are parameters for the RR charge of the D3-branes and distinguish branes
from anti-branes and enter in all the topological terms.
Since the CS forms $\star \omega_3$ will mostly only appear
inside $\hat F_p$ in the following, we will not make this kind of localization
explicit, to keep the notation compact.




\subsection{Equations of motion and constraints}

The most interesting equations of motion to consider are the two sets of
Einstein equations for internal and external indices, the equation
for the dilaton, and the charge conservation constraint coming from the
equation of motion for $C_4$.
Einstein's equations with the general warped
ansatz (\ref{genwarp}) are abbreviated, using the notation of (\ref{estequ})
(see the appendices \ref{appa2} and \ref{appa3}),
\beqn
R_{\mu\nu} &=&
\hat R_{\mu\nu} + \frac12 \hat g_{\mu\nu} \Delta^{1-3b} \hat g^{ij} \hat \nabla_i \left(
  \Delta^{2(b-1)} \partial_j \ln(\Delta) \right) ~=~
\kappa_{10}^2 {\cal S}_{\mu\nu} \ , \non
R_{ij} &=& \kappa_{10}^2 {\cal S}_{ij} \ .
\eeqn
The explicit dependence of $R_{ij}$ on the warp factor is given
in the appendix in (\ref{ricci}). Furthermore, as in the appendix,
$\hat \nabla_i$ is the covariant derivative of $\hat g_{ij}$. \\

The first set of equations allows us to determine the warp factor, which then
has to be checked with the second set. We specialize to maximally symmetric
four-dimensional space-times, i.e.\ $\hat g^{\mu \nu} \hat R_{\mu\nu}=m_4^2$
where $m_4^2$ is zero, positive or negative for Minkowski, de Sitter
or anti-de Sitter space respectively. Then, upon taking the trace, one arrives at
\beqn \label{warpconstraint}
\hat g^{ij} \hat \nabla_i \left(
  \Delta^{2(b-1)} \partial_j \ln(\Delta) \right) &=&
- \frac{\Delta^{-2}}{4} |\hat G_3^{\{0,3\}}|_{\hat g}^2 e^{\Phi} - \frac12 \Delta^{3b-1} m_4^2 \\
&& \hspace{-1cm}
-~ \frac{\Delta^{2+2b}}{2\hat g_4} (\partial_i \alpha)(\partial_j \alpha) \hat g^{ij}
 - \frac{\Delta^{-2} \kappa^2_{10} \mu_3 Q_3}{\sqrt{\hat g_6}}
   \sum_{k=1}^{64} \delta^6(x_k - \bar x_k) \non
&&  \hspace{-1cm}
-~ \frac{\Delta^{-2} \kappa^2_{10} \mu_3 }{\sqrt{\hat g_6}}
  \left( \sum_{\rm branes} \Big(s_m - \frac{i}{3} {\rm tr} (A^i A^j A^k) H_{ijk}\Big) \delta^{6}( x_m - \bar x_m) \right. \non
&&       \left.  +~ e^{\Phi} \Delta^{-2b}
          \sum_{\rm branes} {\rm tr}\, |{\cal F}^{\{0,2\}}|_{\hat g}^2
        \, s_m \delta^{6}( x_m - \bar x_m) \right)\ , \nonumber
\eeqn
where $\hat g_4$ denotes the absolute value of the determinant of $\hat g_{\mu \nu}$ and
the subscript on $|\hat G_3^{\{0,3\}}|_{\hat g}^2$ implies that the contractions are to be performed
with $\hat g^{ij}$. Furthermore, tr$|{\cal F}^{\{0,2\}}|_{\hat g}^2 =
\frac12 \hat g^{ij} \hat g^{kl} f^{abc} f^{ade} A^{b}_i A^{c}_k A^{d}_j A^{e}_l$
is given by $\Delta^4 \frac12 \hat g_{ij} \hat g_{kl} f^{abc} f^{ade} A^{bi} A^{ck} A^{dj} A^{el}$,
when expressed through the metric independent fields $A^{ai}$.
The integral of the right-hand side over the internal space is now forced
to vanish. The D-brane tension and the contributions of the YM sector add up with
the 3-form flux, while the O3-plane tensions serve
as positive contributions on the right-hand side, which may be employed
to balance the negative contributions from the fluxes.\footnote{The
sign conventions are such that negative contributions on the right-hand side
of (\ref{warpconstraint}) are related to positive energy density.} The cosmological
constant and the correction term of the form tr$(A^iA^jA^k)H_{ijk}$ can add to either the positive or negative contributions. \\

Upon specializing to $b=1$ the second set of equations becomes tractable.
Combining the two sets of Einstein equations
one can eliminate the term with the Laplacian acting on the warp factor and obtains
\beqn \label{adee}
\frac{\Delta^{2}}{4} \hat R_{\mu\nu} \hat g^{\mu\nu} +
\frac16 \hat R_{ij} \hat g^{ij} &=&
\frac{1}{3} \hat g^{ij} (\partial_i \ln \Delta )(\partial_j \ln \Delta)
- \frac{\Delta^4}{12 \hat g_4} \hat g^{ij} (\partial_i \alpha) (\partial_j \alpha) \non
&&
-~ \frac{\kappa^2_{10}}{12}
 \left( {\cal T}_{ij} - {\cal T}_{ij}^{\rm 5-flux} \right) \hat g^{ij}\ .
\eeqn
An obvious solution with $\hat R_{\mu\nu}=\hat R_{ij}=0$
consists in $\alpha = \pm \Delta^{-2}$ and $({\cal T}_{ij} -
{\cal T}_{ij}^{\rm 5-flux} ) g^{ij}=0$. This can be satisfied by
${\rm tr}|{\cal F}|^2=0$. There is, however, no constraint on
$\hat G_3$ from (\ref{adee}) since ${\cal T}_{ij}^{\rm 3-flux} g^{ij} = 0$,
cf.\ (\ref{emt}). \\

On the other hand, combining (\ref{warpconstraint}) with the
equation of motion for the external components of $C_4$ leads to a
further constraint. The latter has been discussed in some detail
in section \ref{secaddmod}, where it was pointed out that no
corrections due to the non-abelian CS action occur. It reads
\be \label{bianchi}
(d F_5)^{\{0,6\}} = F_3^{\{0,3\}} \wedge H_3^{\{0,3\}} +
   2\kappa^2_{10} \mu_3 \sum_{\rm branes} q_m \pi_m +
   2\kappa^2_{10} \mu_3 Q_3 \sum_{k=1}^{64} \pi_k \ ,
\ee
and explicitly we have
\beqn \label{bianchi2}
\frac{1}{\sqrt{\hat g_4}} \Big( \hat g^{ij} \hat \nabla_i \partial_j \alpha
+ \frac{4}{\Delta} \hat g^{ij} (\partial_i \Delta)
(\partial_j \alpha) \Big) & = & - \frac{1}{(3!)^2 \Delta} F_{ijk} H_{lmn}
\epsilon^{ijklmn}
\\
&& \hspace{-3cm}
-~
\frac{2\kappa^2_{10} \mu_3}{\sqrt{\hat g_6} \Delta^4} \left(
 \sum_{\rm branes} q_m \delta^6 (x_m - \bar x_m ) +
 Q_3 \sum_{k=1}^{64} \delta^6 (x_k - \bar x_k ) \right)\ . \nonumber
\eeqn
The $\pi_m$ and $\pi_k$ stand for the
delta-function-like 6-forms with support at the location of the 3-branes and 3-planes.
Using the splitting (\ref{splitiasd}) of $\hat G_3$ in (\ref{warpconstraint}) and by adding
$\Delta^2$ times (\ref{warpconstraint}) and $- \frac12 \Delta^4$ times (\ref{bianchi2}),
one can derive the powerful constraint
\beqn \label{thepowerfulconstraint}
&& - \frac{\Delta^4}{2} \hat g^{ij} \hat \nabla_i \partial_j \left(\frac{\alpha}{\sqrt{\hat g_4}} + \Delta^{-2} \right)
 = - \frac{\Delta^6}{2} \hat g^{ij} \partial_i \left( \frac{\alpha}{\sqrt{\hat g_4}} + \Delta^{-2} \right)
\partial_j \left( \frac{\alpha}{\sqrt{\hat g_4}} + \Delta^{-2} \right)
 \non
&& \hspace{2cm} -~ \frac{1}{2} e^{\Phi} {|\hat G_3^{\rm ISD}|_{\hat g}^2}
- \frac{\kappa^2_{10} \mu_3 \Delta^{-2}}{\sqrt{\hat g_6}} e^{\Phi}
   \sum_{\rm branes} {\rm tr}\, |{\cal F}^{\{0,2\}}|_{\hat g}^2
   \, s_m \delta^{6}( x_m - \bar x_m)
\non
&& \hspace{2cm} -~
\frac{\kappa^2_{10} \mu_3}{\sqrt{\hat g_6}}
 \sum_{\rm branes} \left( s_m - q_m \right) \delta^6 (x_m - \bar x_m ) - \frac12 \Delta^4 m_4^2\ .
\eeqn
This generalizes the result of \cite{hepth0105097} by including the world-volume fields of the
D3-branes and allowing for $m_4^2 \not =0$.
Integrating the total derivative, a number of restrictions follow. The four-dimensional
cosmological constant has to satisfy $m_4^2\le 0$, excluding de Sitter space within
the chosen ansatz. In addition, $s_m = q_m =1$ follows, so there are no anti-branes.
For vanishing cosmological constant, i.e.\ $\hat g_{\mu \nu} = \eta_{\mu \nu}$,
all terms on the right-hand side are negative semi-definite and therefore have
to vanish individually. Thus,
\beqn \label{solution}
\alpha = - \Delta^{-2}\ ,\quad
|\hat G_3^{\rm ISD}|^2 = {\rm tr}\, |{\cal F}^{\{0,2\}}|^2 = 0 \ .
\eeqn
These conditions are
equivalent to asking for a global minimum of the effective potential (\ref{effpot})
obtained by neglecting the warp factor
\be
{\cal V}_{\rm eff} = 0 ~~~~\Longleftrightarrow~~~
\left( \hat G_3^{\rm ISD}=0\ {\rm and}\ {\rm tr}|{\cal F}|^2=0 \right) \ .
\ee
Since this potential is non-negative, global minima are precisely the Minkowski vacua.
These conclusions are clearly not valid for four-dimensional anti-de Sitter space, which
seems to demand a choice for $\alpha$ such that $\alpha/\sqrt{\hat g_4}$ is
independent of the external coordinates.
\\

Using the formulas from appendix \ref{appa2} and \ref{appa3} it is possible
to show that the conditions (\ref{solution}) ensure that the
full set of Einstein equations are satisfied if
\beqn \label{warp}
\hat g^{ij} \hat \nabla_i \partial_j \Delta^2 & = & -\frac12 e^\Phi |\hat G_3|^2_{\hat g} \\
&& \hspace{-2.5cm} -~ \frac{2\kappa^2_{10} \mu_3}{\sqrt{\hat g_6}} \left(
 \sum_{\rm branes} \Big(s_m - \frac{i}{3} {\rm tr} (A^i A^j A^k) H_{ijk}\Big) \delta^6 (x_m - \bar x_m )
+ Q_3 \sum_{k=1}^{64} \delta^6 (x_k - \bar x_k ) \right)\ . \nonumber
\eeqn
To show this one has to use the fact that ${\cal T}_{ij}^{\rm 3-flux}=0$ for purely
IASD (or ISD) $\hat G_3$, as was noticed in \cite{hepth0009211}. Imposing
(\ref{solution}), the equation determining the warp factor (\ref{warp}) is
actually equivalent to (\ref{bianchi2}). \\

The other equations of motions are not too difficult to find.
The equation for $\Phi$ is
\beqn
\frac{\Delta^3}{\sqrt{-g}}
\partial_I \left( g^{IJ} \sqrt{- g} \partial_J \Phi \right) &=&
    \frac12 \left( e^{\Phi} |\hat F_3^{(0,3)}|_{\hat g}^2 - e^{-\Phi} |H_3^{(0,3)}|_{\hat g}^2 \right) \\
&& 
+~ \frac{\kappa^2_{10} \mu_3}{\sqrt{\hat
g_6}\Delta^2} e^{\Phi}
    \sum_{\rm branes} {\rm tr}\,
 |{\cal F}^{\{0,2\}}|_{\hat g}^2 \, s_m \delta^{6}( x_m - \bar x_m )\ . \nonumber
\eeqn
This equation imposes a
vanishing of the dilaton tadpole. By four-dimensional
Poin\-ca\-r\'e invariance
the right-hand side of the equation has to integrate to zero.
An IASD 3-form fulfills $|\hat F_3|^2 = e^{-2\Phi}|H_3|^2$ and therefore the YM potential
must vanish, tr$|{\cal F}|^2=0$. Thus, the solutions of the constraint
coming from the Einstein equations are compatible with the vanishing of the dilaton 
tadpole.
\\

To summarize the logic: Asking for a Minkowski
vacuum leads to (\ref{warpconstraint}) with $m^2_4 = 0$, combining with
(\ref{bianchi}) implies (\ref{solution}) and is compatible with
$\hat R_{ij} = 0$ at constant dilaton.
We are then dealing with a warped product of a Calabi-Yau or torus and
Minkowski space, where the warp factor is determined via (\ref{warp}), which is equivalent to
(\ref{bianchi2}).
The situation described above is a variant of the known no-go
theorems of \cite{deWit:1986xg,Maldacena:2000mw,Ivanov:2000fg,Gauntlett:2002sc}. Several escape
routes are at least in principle known: In an anti-de Sitter vacuum, a positive term would
appear on the right-hand side of (\ref{warpconstraint}) that could balance
the vacuum energy of the potential terms. If higher curvature corrections to
the action were considered the semi-definitness could also be spoiled
and de Sitter solutions may exist as well.
Perturbative \cite{Becker:2002nn} or non-perturbative \cite{Kachru:2003aw} corrections 
to the effective potential also appear to modify the conclusions drawn above.
Finally, one may also want to allow an explicitly
time-dependent background \cite{Townsend:2003fx}. \\


\subsection{Large volume scaling limit: separation of mass scales}

The solution given by the conditions (\ref{solution}) is too simple to fix all
the metric moduli; e.g.\ it always leaves the overall volume of the
internal space as a free parameter. We will later come back to the
possibility to fix also this by adding world-volume gauge fluxes on higher-dimensional branes.
For the moment, one can observe that, like in \cite{hepth0105097},
the relations that determine the solution are all invariant under constant
rescaling of the metric. This is in accord with the form of the effective
potential despite the fact that both terms in (\ref{effpot}) scale
differently under $g_{ij} \rightarrow t g_{ij}$. The reason is
that for a Minkowski vacuum both terms have to
vanish separately. This agrees with the results of \cite{hepth0206241,D'Auria:2003jk}
where the no-scale structure of the gauged four-dimensional ${\cal N}=4$ supergravity theory, that
is the effective description of the present scenario, was
demonstrated explicitly. \\

While the scale-independence of (\ref{solution}) is unfortunate in that it does not
lead to a fixing of the volume modulus, it has been argued to allow for a limit of
parametrically large volume where the warping becomes insignificant
\cite{hepth0105097,hepth0201029}. In this limit, one can perform a standard dimensional reduction
of the action to four dimensions. Then (\ref{effpot}) really takes the
role of an effective potential of a theory obtained by expanding around a given
solution to the equations of motion.
This is justified by inspection of (\ref{bianchi2}) (or alternatively of
(\ref{warpconstraint})): The metric factors on the
left-hand side of (\ref{bianchi2}) scale like $t^{-1}$, on the right-hand side like $t^{-3}$. As
$F_{ijk}$ and $H_{ijk}$ are just constants, the warp factor itself has to go as
$1+t^{-2}$ at large $t$. This argument is, of course, only valid away
from the positions of the branes and planes. Ignoring the contributions from
these regions one can then set $\Delta$ to a constant in the large $t$ limit and
derive the effective potential (\ref{effpot}) by dimensional reduction.
In this situation, it appears challenging to find ways to
stabilize the volume modulus in the effective theory that is only valid at very large
volume. Given a correction to (\ref{effpot}) that fixes the volume,
one would need to make sure that this is done at large enough values that
the effects of the warping are still negligible.
In principle, it would of course be much nicer if one
could do a reduction without using the scaling argument by explicitly
including the warp factor, similar to \cite{DWG}. \\

The scaling argument can also be used for branes of higher dimensions.
In order to neglect the warp factor at large $t$ the contributions of all
relevant types of matter on the right-hand side of the Einstein equations
have to fall off faster than $t^{-1}$. The tension of an arbitrary D$p$-brane
(in Einstein-frame) leads to
\be
{\cal S}^{{\rm D}p}_{\mu\nu} = \frac{p-7}{8}
 \frac{\mu_p e^{(p-3)\Phi /4}}{\sqrt{g_{\perp}}}
 g_{\mu\nu} \delta^{9-p}(x_m - \bar x_m) \ .
\ee
Here $g_{\perp}$ indicates the determinant of the metric
restricted to the normal
bundle of the brane. If the metric is constant this is the transverse volume.
${\cal S}_{\mu\nu}$ then scales like $t^{(p-9)/2}$ and
$p < 7$ is required for the scaling argument to work. \\

The validity of the approach in addition requires a separation of
the flux-induced masses and the masses of KK modes that have been
neglected throughout. It has been argued in \cite{KM,hepth0201028}
that the ratio of masses induced by the 3-form flux and the KK
masses scales like
\be \label{scales}
(m_{\rm 3-flux} \, : \,m_{\rm KK}) = (R^{-3} \, : \, R^{-1})\ ,
\ee
where $R$ denotes the dimensionless (average) radius of the background
torus. Thus also from here we see that we need a large volume, this time to ensure
a decoupling of the KK states. The same reasoning is still
true in the case at hand since the additional second term in the potential
(\ref{effpot}) does not introduce any new mass terms for the geometric
moduli. This is due to the fact that ${\cal F}^{\{0,2\}}$ vanishes in the
background, independently of the geometric moduli. Thus the masses only depend
on the 3-form flux, as is also apparent from the mass formulas in
\cite{D'Auria:2003jk}. This would be different if one considered higher-dimensional
branes with internal world-volume fluxes, to which we come back in the next section.
In that case one would expect an additional mass scale from the world-volume fluxes,
similar to the situation in the heterotic string discussed in \cite{KM} and an intermediate
scale appears, below the KK but still above the 3-form flux scale:
\be \label{scales2}
(m_{\rm 3-flux} \, : \,m_{\rm 2-flux}: \,m_{\rm KK}) = (R^{-3} \, : \, R^{-2}: \, R^{-1})\ .
\ee
%


\section{Generalization of the open string sector}
\label{sec6}

In this final section we want to go beyond the well-defined
framework of the model obtained from type I by T-duality, and allow
higher-dimensional D-branes of even dimension and with internal
world-volume gauge fluxes. This means we use
\beqn \label{fstrflux}
{\cal F}^a_{ij} = f^a_{ij} + f^{abc}  A^b_i A^c_j \ ,
\eeqn
instead of (\ref{fmnexplicit}).
The constant flux parameters $f_{ij} = f_{ij}^a T^a$ take values in the
Cartan subalgebra of the gauge group.
They characterize a non-trivial gauge bundle on the world-volume.
It is well known that the open string boundary conditions
with constant fluxes change from Dirichlet to mixed
Dirichlet-Neumann conditions.
This implies that performing two T-dualities on a D$p$-brane wrapping a
two-dimensional torus with flux $f_{12} \not = 0$ turns it into another D$p$-brane of
identical dimension, but with the flux inverted. Therefore, introducing
such gauge fluxes into the original type I string theory, assuming a factorization of the
background into $\mathbb{T}^6=(\mathbb{T}^2)^3$ for simplicity, could have
taken us to a T-dual theory with D3, D5, D7 or D9-branes (depending on the number of
internal directions with world-volume flux) wrapping 0, 2, 4 or all
directions of the $\mathbb{T}^6$ after the duality,
and with non-trivial gauge fluxes on their world-volume.
At the same time, the O9-planes would still map to O3-planes.
The D5, D7 and D9-brane charges then have to cancel among the
branes, so that some of them need to carry negative charge as well
and behave as anti-branes. This leads to the conclusion that there
are no supersymmetric vacua with non-trivial world-volume flux in
a flat background on a torus, which is in accord with the known
fact that the toroidal type I compactification with world-volume
gauge fluxes on D9-branes, that has been of some phenomenological
interest recently, does not have a supersymmetric ground state
\cite{Blumenhagen:2001te}. To achieve a
better behaved model one may consider orbifold compactifications
which also have O7-planes in addition to the O3-planes. In this
case supersymmetric configurations of D9-branes with world-volume
fluxes do exist
\cite{Blumenhagen:2000ea,Cvetic:2001nr,Blumenhagen:2002gw,Honecker:2003vq}.\footnote{This
type of models, combined with 3-form fluxes, has been considered
recently in \cite{Blumenhagen:2003vr,Cascales:2003zp}.} Here we
content ourselves with studying a few general properties of
D9-brane flux vacua to obtain a rough impression of the effects of
warping and perspectives for moduli
stabilization. \\

In the presence of internal gauge fluxes, we have to use the DBI
effective action including terms with higher powers of ${\cal F}$,
since in the presence of world-volume fluxes the higher powers in
${\cal F}$ contribute to leading order in a derivative expansion
of the effective action. Furthermore, some of the CS corrections
in $\omega_7$ are no longer higher order and need to be
incorporated into the action as in (\ref{rrsymm}). On the other
hand, the full DBI effective action can only reliably be used for
an abelian gauge group. Therefore we now pass to a $U(1)^N$ gauge
group by turning on Higgs fields. In this way we lose the 
possibility to produce chiral matter in various
representations, which is one of the main motivations to be
interested in this variant of the model
\cite{Blumenhagen:2000wh,Angelantonj:2000hi,spanier}. However, as said above, here we are
mainly interested in the prospects for moduli stabilization. To
simplify the notation we also restrict ourselves to D9-branes,
using
\beqn \label{dbib}
{\cal S}_{\rm DBI} &=& - \mu_9 \int_{\mathbb{R}^4\times {\cal M}_6}{d^{10} x\
                             \sqrt{-g_4} e^{3\Phi/2}
\sqrt{ {\rm det} ( g_{ij} + {\cal F}_{ij} e^{-\Phi/2}) }} \non
&=& - \int_{\mathbb{R}^4}{d^4 x\ \sqrt{-g_4}\
{\cal V}_{\rm DBI}}
\eeqn
as effective world-volume action. Considering D5- or D7-branes should pose no further
difficulties and would lead to similar conclusions. \\


\subsection{Supersymmetry and $\kappa$-symmetry}

In this section we discuss the conditions for preserving world-volume supersymmetry
on D9-branes in the simultaneous presence of 3-form fluxes and O3-planes
in the background.
In their absence they are
given by the calibration conditions found in \cite{hepth9911206}. \\

The background bulk fields are not only
required to satisfy the conditions (\ref{solution})
in order to solve the equations of motion, but we also impose $G_3^{\rm IASD}$ to be of
type $(2,1)$ and primitive \cite{hepth0009211,hepth0010010} to preserve supersymmetry.
One then finds that in this background there is a Killing spinor
that can be expressed as a covariantly constant spinor rescaled by a
scalar function only \cite{hepth0009211},
\beqn \label{killing}
\epsilon = \Delta^{-1/4} \eta \ .
\eeqn
%
The background is 1/2 BPS, since $\eta$ also satisfies a chirality
projection.
In general, world-volume supersymmetry of D-branes in non-trivial backgrounds is
preserved if the supervariations can be absorbed by $\kappa$-symmetry
variations. This is translated into the projection
\beqn \label{kappa}
\Gamma \epsilon = \epsilon \ ,
\eeqn
where $\Gamma=\Gamma({\rm P}[g],{\cal F})$ is the $\kappa$-projector, an operator
depending on the pull-back of the (warped) metric and the gauge field strength
${\cal F}$, and $\epsilon$ is a Killing spinor of the background bulk theory.
Explicitly, the $\kappa$-symmetry projector is defined by
\beqn
\Gamma = e^{-a/2} \left(  \Gamma_{0...p} \otimes i\sigma_2 \right) e^{a/2}
\eeqn
with
\beqn
a &=& \frac{1}{2} Y_{ij}
      E^i_A E^j_B \Gamma^{AB} \otimes \sigma_3
   = \frac{1}{2} \Delta Y_{ij}
      \hat E^i_A \hat E^j_B \Gamma^{AB} \otimes \sigma_3 \ , \non
Y_{ij} &=& {\rm ``arctan{\mbox{''}}}({\cal F}_{ij}) \ , \quad
\hat g_{ij} ~=~ \delta_{AB} \hat E_i^A \hat E_j^B \ ,
\eeqn
where ${\rm ``arctan}$'' is explained in \cite{hepth9705040}, to
which we refer the reader for further information on the notation
and techniques. The details do not play a prominent role in our
discussion here. The important point for us is the existence of a
Killing spinor of the type (\ref{killing}) in a background of
supersymmetric 3-form fluxes and O3-planes, a rescaled
(covariantly) constant Killing spinor. Neglecting the backreaction
of the D9-branes themselves, one can use this Killing spinor to
derive analogous calibration conditions as found in
\cite{hepth9911206} by just modifying the background metric via
inclusion of the warp factor. A simple example with a flat
background $\hat g_{ij} = \delta_{ij}$ has been studied in
\cite{hepth0203019}.
\\


\subsection{Calibrated branes}

We are now ready to apply the results of \cite{hepth9911206} to find the
supersymmetry conditions for D9-branes with world-volume gauge fluxes in a given background of
O3-planes and 3-form fluxes.
It is then required that the purely holomorphic components of ${\cal F}$ vanish and that
\beqn \label{mmms}
e^{i\theta} \sqrt{{\rm det}( g_{ij} + {\cal F}_{ij} ) }\,
dx_1\wedge\, ... \wedge dx_6
&=&
\frac{1}{3!} (J+i{\cal F})^3 \ , \\
\cos(\theta) \left( \frac{1}{2!}  J \wedge J \wedge {\cal F} - \frac{1}{3!}
{\cal F} \wedge {\cal F} \wedge{\cal F} \right)
&=&
\sin(\theta) \left( \frac{1}{3!} J\wedge J\wedge J - \frac{1}{2!}
{\cal F}\wedge {\cal F}\wedge J \right)
\nonumber
\eeqn
hold for some phase $\theta$ on all the D9-branes. $J = \Delta \hat J$ now stands
for a rescaling of the K\"ahler form $\hat J$ of the Calabi-Yau.
For non-trivial $\Delta$ this $J$ is not closed.
The background O3-planes (and any possible O7-planes) are calibrated with an angle $\theta=3\pi/2$.
If one neglects the warp factor, i.e.\ uses $\Delta = 1$,
an overall world-volume supersymmetry can be preserved by adopting the same angle
for all the D9-branes as well.
It is important to note that fixing $\theta$, induces an implicit
dependence of the K\"ahler moduli on ${\cal F}$.
The scalar potential for the K\"ahler moduli in the Einstein frame reads
\beqn \label{potent}
{\cal V}_{\rm DBI} \sim
\int_{{\cal M}_6} \sqrt{ {\rm det} ( g_{ij} + {\cal F}_{ij} e^{-\Phi/2}) } =
\int_{{\cal M}_6} {\rm Re} \Big( \frac{e^{-i \theta}}{3!} (J+i{\cal F}e^{-\Phi/2} )^3 \Big)\ .
\eeqn
One further needs to impose the charge conservation constraint,
which now includes the open string CS interactions
$\mu_9 C_4 \wedge {\rm ch}({\cal F})$, such that (\ref{bianchi}) is modified to
\beqn \label{tad} \nonumber
d F_5 = F_3 \wedge H_3 +
   2\kappa^2_{10} \mu_3 Q_3 \sum_{k=1}^{64} \pi_k
  +2\kappa^2_{10} \frac{\mu_9}{3!} \sum_{\rm branes} {\cal F} \wedge {\cal F} \wedge {\cal F}\ ,
\eeqn
which leads to the condition
\beqn \label{tadpole}
\frac{1}{3 (4\pi^2 \alpha')^3}
\sum_{\rm branes} \int_{{\cal M}_6} {\cal F} \wedge {\cal F} \wedge {\cal F}
= 32 - N_{\rm flux} \ .
\eeqn

Similarly, the couplings of RR 8-form potentials in the open string
CS action leads to a tadpole constraint involving ${\cal F}$,
which states that the sums of the effective RR 7-brane charge has to be zero.
As already said above, in order to fulfill (\ref{mmms}) for all branes with the same angle
$\theta$ one would have to depart from the
flat torus example or to include O7-planes via an orbifold construction as in
\cite{Blumenhagen:2003vr,Cascales:2003zp}.
Then (\ref{mmms}) together with $G_3^{\rm ISD}=0$
ensures the vanishing of the total potential energy and (\ref{mmms}) degenerates into
\beqn \label{fixkaehler}
J \wedge  J \wedge  J
= 3  J \wedge {\cal F} \wedge {\cal F}
\eeqn
for any individual stack of branes.
This constraint ensures the balancing of the 9-brane and 5-brane tensions, while 3-
and 7-brane charges and tensions are unconstrained and only cancel upon adding
the O-planes. It
fixes the overall radius in an obvious manner.
The natural scale for its value is however of the order of $\sqrt{\alpha'}$, with a
constant of proportionality given by a combination of flux numbers.
As one should maintain the splitting of scales (\ref{scales2}) it would be
required to have hierarchically large flux quantum numbers, which by
(\ref{tadpole}) would demand some ``integer fine tuning'' to cancel the
tadpoles. Furthermore, once the volume modulus gets fixed at a finite value,
the warp factor is not trivial anymore.
Writing $J = \Delta \hat J$ for closed $\hat J$ one sees that
(\ref{fixkaehler}) can not be fulfilled anymore. It would actually be equivalent to
$\hat J^3=\Delta^{-2} \hat J \wedge {\cal F}^2$, where the left-hand side is closed and
the right-hand side not. \\

This probably does not mean that the volume modulus cannot be stabilized in this way, since
the qualitative stabilization mechanism due to the balancing of the 9-brane and 5-brane tensions
should still take place. On the other hand, it seems that the non-trivial warp factor leads to a
breaking of supersymmetry, once the volume is fixed. A similar result
has been found in \cite{hepth0203019}. However, for a
complete analysis one would also have to take into account the
backreaction of the background towards the presence of the D9-branes and possible
O7-planes. This would probably lead to a significant modification of the calibration
condition (\ref{mmms}) but one should expect the above results on the stabilization of the volume
to be qualitatively true also in the full story \cite{Blumenhagen:2003vr,Cascales:2003zp}.
\\


\begin{center}
{\bf Acknowledgements}
\end{center}

It is a great pleasure to acknowledge stimulating discussions and helpful explanations
provided to us by Katrin and Melanie Becker, Massimo Bianchi,
Fawad Hassan, Shamit Kachru, Jurg K\"appeli, Jan Louis, Gianfranco Pradisi,
Augusto Sagnotti, Henning Samtleben, Michael Schulz, Mario Trigiante and Marco Zagermann.
During the work B.~K.\ enjoyed the hospitality of the Universit\`a di Roma ``Tor Vergata'',
while M.~B.\ and M.~H.\ performed part of this work at the University of Utrecht and the Spinoza Institute.
M.~H.\ would also like to thank the theory group at CERN for hospitality during the
final stages of the work.
We further like to thank the organizers of the RTN network conferences {\it Superstring
Theory} at Cambridge 2002, where this work was initiated, and those of
{\it The quantum structure of spacetime and the geometric nature of fundamental interactions} at
Torino 2003. The work of B.~K.\ was supported by the German Science Foundation (DFG) and in part by
funds provided by the U.S. Department of Energy (D.O.E.) under cooperative research agreement
$\#$DF-FC02-94ER40818.
The work of M.~B.\ and M.~H.\ was supported in part by I.N.F.N., by the
E.C. RTN programs HPRN-CT-2000-00122 and HPRN-CT-2000-00148, by the
INTAS contract 99-1-590, by the MURST-COFIN contract 2001-025492 and
by the NATO contract PST.CLG.978785.


\newpage
\begin{appendix}

\section{Technicalities}

\subsection{Conventions and notation}
\label{conventions}

Tangent-frame indices are generally denoted $I,J,\, ... = 0,\, ...\, , 9$, which
decompose into
$i,j,\, ... = 4,\, ...\, , 9$ and $\mu,\nu,\, ... = 0,\, ...\, , 3$.
Once we complexify coordinates, we use $m,n,\, ...$.
We use the following standard conventions for differential forms:
For $\Omega_p \in \bigwedge^p T^* {\cal M}_{10}$,
${\cal M}_{10} = \mathbb{R}^4 \times \mathbb{T}^6$, write
\beqn \label{defforms}
\Omega_p &=& \frac{1}{p!} \Omega_{I_1\, ... \, I_p}
 dx^{I_1} \wedge\, \cdots\, \wedge dx^{I_p}\ , \non
|\Omega_p|^2 &=& \frac{1}{p!} \Omega_{I_1\, ... \, I_p}
   \bar\Omega^{I_1\, ... \, I_p} \ ,
\eeqn
where complex forms obviously refer to the complexified cotangent bundle,
for $\Omega^{\{n,p-n\}}_p \in \bigwedge^n
T^* \mathbb{R}^4 \times \bigwedge^{p-n} T^*\mathbb{T}^6$ analogously
\beqn
\Omega^{\{n,p-n\}}_p &=&
 \frac{1}{n!(p-n)!} \Omega_{\mu_1\, ... \, \mu_n i_1\, ...\, i_{p-n}}
dx^{\mu_1} \wedge\, \cdots\, \wedge dx^{\mu_n} \wedge
dx^{i_1} \wedge\, \cdots\, \wedge dx^{i_{p-n}}\ , \non
|\Omega^{\{n,p-n\}}_p|^2 &=& \frac{1}{n!(p-n)!}
\Omega_{\mu_1\, ... \, \mu_n i_1\, ... \, i_{p-n}}
   \bar\Omega^{\mu_1\, ... \, \mu_n i_1\, ... \, i_{p-n}}\ .
\eeqn
Ten-dimensional Hodge duality is defined by
\be
* \Omega_p = \frac{1}{p!(10-p)!} {\epsilon^{I_1\, ...\, I_{p}}}_{I_{p+1}\, ...\, I_{10}}
\Omega_{I_{1}\, ...\, I_{p}}
dx^{I_{p+1}} \wedge\, \cdots\, \wedge dx^{I_{10}}\ ,
\ee
and the six-dimensional Hodge star is given by
\be \label{hodgesix}
(\star \Omega_p)^{\{n,6-p+n\}}_{\mu_1 \ldots \mu_n j_1 \ldots j_{6-p+n}}
= \frac{1}{(p-n)!} {\epsilon^{i_1 \ldots i_{p-n}}}_{j_1 \ldots j_{6-p+n}}
\Omega_{\mu_1\, ...\, \mu_n i_{1}\, ...\, i_{p-n}}\ .
\ee
We also define derivative operators
\beqn \label{dproj}
d^{\{1,0\}} \Omega^{\{n,p-n\}}_p &=& (d\Omega^{\{n,p-n\}}_p)^{\{n+1,p-n\}} \ , \non
d^{\{0,1\}} \Omega^{\{n,p-n\}}_p &=& (d\Omega^{\{n,p-n\}}_p)^{\{n,p+1-n\}} \ .
\eeqn
For the totally antisymmetric tensor $\epsilon$ we use the convention
\be
\epsilon_{1\, ... \, D} = \pm \sqrt{g_D}\ ,\quad
\epsilon^{1\, ... \, D} = \frac{1}{\sqrt{g_D}}\ ,
\ee
with $g_D$ being the (absolute value of the) determinant of the metric
and the sign depends on the signature of the metric.


\subsection{Einstein equations with warp factors}
\label{appa2}

In this appendix we give the results for the Ricci tensor of
a general warped metric ansatz (\ref{genwarp}).
With that ansatz the Christoffel symbols are found
\beqn
&& \Gamma^\mu_{\nu\rho} = \hat \Gamma^\mu_{\nu\rho}\ , \quad
\Gamma^\mu_{ij} = \Gamma^i_{\mu j} = 0\ , \non
&& \Gamma^\mu_{\nu i} =
   -\frac{1}{2} \delta^\mu_\nu \partial_i \ln (\Delta)\ , \quad
\Gamma^i_{\mu\nu} =
   \frac12 \hat g_{\mu\nu} \Delta^{-1-b} \hat g^{ij} \partial_j \ln (\Delta) \ , \non
&& \Gamma^i_{jk} = \hat \Gamma^i_{jk} +
   \frac{b}{2} \left( \delta^i_j \partial_k \ln (\Delta) +
                      \delta^i_k \partial_j \ln (\Delta) -
                      \hat g_{jk} \hat g^{il} \partial_l \ln (\Delta) \right) \ .
\eeqn
%
This leads to
\beqn \label{ricci}
R_{\mu\nu}
&=& \hat R_{\mu\nu} +
   \frac12 \hat g_{\mu\nu} \Delta^{-1-b} \left( \hat g^{ij} \hat \nabla_i \partial_j \ln (\Delta) -
   2 (1-b)  \hat g^{ij} \partial_i \ln (\Delta) \partial_j \ln (\Delta)  \right)  \non
&=& \hat R_{\mu\nu} +
   \frac12 \hat g_{\mu\nu} \Delta^{1-3b}\, \hat g^{ij} \hat \nabla_i \left(
   \Delta^{2(b-1)} \partial_j \ln (\Delta) \right) \ , \non
R_{ij} &=& \hat R_{ij} + 2(1-b) \hat \nabla_i \partial_j \ln (\Delta)
   + (b^2 -2b -1) (\partial_i \ln (\Delta))(\partial_j \ln (\Delta)) \non
&&
   + b(1-b) \hat g_{ij} \hat g^{kl} \partial_k \ln (\Delta) \partial_l \ln (\Delta) -
   \frac{b}{2} \hat g_{ij} \hat g^{kl}\hat \nabla_k \partial_l \ln(\Delta) \ ,
\eeqn
where
$\hat \nabla_i$ involves the
Christoffel symbols $\hat \Gamma^i_{jk}$.
We define the energy momentum tensor for some action ${\cal S}[g]$ by
\be
{\cal T}_{IJ} =
  -\frac{2}{\sqrt{-g}} \frac{\delta {\cal S}}{\delta g^{IJ}} \ ,
\ee
${\cal T}$ denoting its trace ${\cal T}_{IJ} g^{IJ}$,
and the Einstein equations are then written
\be  \label{estequ}
\frac{1}{\kappa^2_{10}}
R_{IJ} = {\cal S}_{IJ} ,\quad
{\cal S}_{IJ} = {\cal T}_{IJ} - \frac18 g_{IJ} {\cal T} \ .
\ee
In a flat Minkowski vacuum ($\hat R_{\mu \nu} = 0$) the trace of the
space-time components then always takes the form
\be
\frac{1}{\kappa^2_{10}} \hat g^{ij} \hat \nabla_i \left(
   \Delta^{2(b-1)} \partial_j \ln (\Delta) \right) =
\frac14 \Delta^{3b-2} \left( {\cal T}_{\mu\nu} g^{\mu\nu} -
                  {\cal T}_{ij} g^{ij} \right) \ .
\ee
From this follows the famous constraint \cite{deWit:1986xg} that the right-hand side,
with the given warped ansatz and without higher derivative
terms in the action, has to integrate to zero.
Thus, whenever it is positive or negative definite, it has to vanish,
which then implies the absence of fluxes and warp factors. \\


\subsection{Energy momentum tensors}
\label{appa3}

In this section we give the explicit forms of the energy momentum tensors
of the various background contributions referring to the ten-dimensional
action (\ref{eini}) in the Einstein-frame.
The only Poincar\'e invariant background form fields we turn on are
the internal components of the 3-form, i.e.\ the fluxes $\hat G_3^{\{0,3\}}$, and the
components $F_5^{\{4,1\}}$ and $F_5^{\{0,5\}}$ (i.e.\ we do not consider
any background for the RR axion that would be needed in discussing
D7-brane solutions, cf.\ \cite{GP}).
The 5-form ansatz has to be self-dual and was given in (\ref{5-formansatz}).
It satisfies $F_5 = (1+*) d\alpha \wedge dx^0 \wedge\, ... \,
\wedge dx^3$,
such that we have in particular $|F_5|^2 = 0$.
For these fluxes one finds the following
contributions to the energy momentum tensor,
\beqn \label{emt}
{\cal T}^{\rm 3-flux}_{\mu\nu} &=&
- \frac{\Delta^{-3b}}{4\kappa^2_{10}}
  e^\Phi g_{\mu\nu} |\hat G_3^{\{0,3\}}|^2_{\hat g} \ , \\
{\cal T}^{\rm 3-flux}_{ij} &=& - \frac{1}{4\kappa^2_{10}} e^\Phi \left(
  g_{ij} \Delta^{-3b} |\hat G_3^{\{0,3\}}|^2_{\hat g} -
   (\hat G_3^{\{0,3\}})_{(i |kl|}
            (\hat{\bar{G}}_3^{\{0,3\}})_{j)mn}\, g^{km} g^{ln} \right) \ ,
\non
{\cal T}^{\rm 5-flux}_{\mu\nu} &=& \frac{1}{4\kappa^2_{10}} \left(
   \frac{1}{3!} (F_5)_{\mu \rho\sigma\tau i}
                (F_5)_{\nu \alpha\beta\gamma j}\,
   g^{\rho\alpha} g^{\sigma\beta} g^{\tau\gamma} g^{ij}
  \right)
\non
&=& - \frac{1}{4\kappa^2_{10}g_4}
 g_{\mu\nu} (\partial_i \alpha)(\partial_j \alpha) g^{ij} \ ,
\non
{\cal T}^{\rm 5-flux}_{ij} &=& \frac{1}{4\kappa^2_{10}} \left(
   \frac{1}{4!} (F_5)_{iKLMN} (F_5)_{jOPQR}\,
    g^{KO} g^{LP} g^{MQ} g^{NR}  \right)
\non
&=& \frac{1}{2\kappa^2_{10}g_4}
 \left( \frac12 g_{ij} (\partial_k \alpha)(\partial_l \alpha)  g^{kl}
  - (\partial_i\alpha) (\partial_j \alpha)  \right) \ .
\nonumber
\eeqn
%
Note ${\cal T}^{\rm 3-flux}_{ij} g^{ij}=
{\cal T}^{\rm 5-flux}_{IJ} g^{IJ}=0$.
If one specializes to $b=1$, $\alpha = - \Delta^{-2}$ and $\hat g_{\mu \nu} = \eta_{\mu \nu}$
one can rewrite (\ref{ricci}) into \cite{hepth0009211}
\beqn \label{ricci5form}
 R_{\mu\nu} - \kappa^2_{10} {\cal T}^{\rm 5-flux}_{\mu\nu} &=&
  \frac14 \eta_{\mu\nu} \Delta^{-4} \hat g^{ij} \hat \nabla_i
    \partial_j (\Delta^2) \ ,
\non
 R_{ij} - \kappa^2_{10} {\cal T}^{\rm 5-flux}_{ij} &=&
  \hat R_{ij} - \frac14 \hat g_{ij} \Delta^{-2} \hat g^{kl} \hat \nabla_k
    \partial_l (\Delta^2) \ .
\eeqn
This is the starting point for the self-dual 3-brane solution.
In the background the only allowed components of the open string gauge fields are ${\cal F}^{\{0,2\}}$.
From (\ref{d3local}) we then get for the open string sector
\beqn
{\cal T}^{\rm op}_{IJ} &=& {\cal T}^{\rm ten}_{IJ}
 + {\cal T}^{\rm YM}_{IJ}, \non
{\cal T}^{\rm ten}_{\mu\nu} &=& - \frac{\mu_3}{\sqrt{g_6}}
   g_{\mu\nu} \sum_{\rm branes} \Big(s_m - \frac{i}{3} {\rm tr} (A^i A^j A^k) H_{ijk}\Big)
\delta^{6} (x_m - \bar x_m), \quad
{\cal T}^{\rm ten}_{ij} ~=~ 0
   , \non
{\cal T}^{\rm YM}_{\mu\nu} &=& - \frac{\mu_3\Delta^{-2b}}{2\sqrt{g_6}}
   g_{\mu\nu} e^{\Phi}
  \sum_{\rm branes} {\rm tr}|{\cal F}^{\{0,2\}}|_{\hat g}^2 \, s_m \delta^{6} (x_m - \bar x_m)\ , \non
{\cal T}^{\rm YM}_{ij} &=& \frac{\mu_3}{\sqrt{g_6}}
  e^{\Phi}  g^{kl}
  \sum_{\rm branes} {\rm tr} ({\cal F}_{ik} {\cal F}_{jl}) \, s_m \delta^{6} (x_m - \bar x_m)\ ,
\eeqn
where we added the correction from the non-abelian DBI action to
the tension. For orientifold planes we write
\be
{\cal T}^{{\rm O}3}_{\mu\nu} =
  - \frac{\mu_3 Q_3 }{\sqrt{{g}_6}}  g_{\mu\nu}
 \sum_{k=1}^{64} \delta^{6}( x_k - \bar x_k ) , \quad
{\cal T}^{{\rm O}3}_{ij} = 0 \ .
\ee
One then also has ${\cal T}^{\rm YM}_{IJ} g^{IJ}=0$.


\subsection{T-duality of RR forms}

In this section we specialize the rules for dualizing RR forms
given in \cite{hepth9907132} to the case relevant in this paper.
The essential step is to replace a RR $p$-form $\Omega_p$ with a state
\be \label{rroperator}
\Omega_p^{\{q, p-q\}} \rightarrow  \frac{1}{(p-q)!}
 \Bigg( \frac{1}{q!} \Omega^{\{q,p-q\}}_{\mu_1 \ldots \mu_q i_1\, ... \, i_{p-q}}
 dx^{\mu_1} \wedge \, \cdots \, \wedge dx^{\mu_q} \Bigg)
 \psi^{i_1 \dag}\, \ldots\, \psi^{i_{p-q} \dag} \vert 0 \rangle\ ,
\ee
where $\psi^{i \dag}$ and $\psi_{i}$ are raising and lowering operators
\be
\{ \psi^{i \dag}, \psi_{j} \} = \delta^i_j\ , \quad
\{ \psi_{i}, \psi_{j} \} = 0 = \{ \psi^{i \dag}, \psi^{j \dag} \}\ ,
\ee
acting on a vacuum with $\psi_i \vert 0 \rangle = 0$ and
$\langle 0 \vert 0 \rangle = 1$.
The elements of the duality group, $\Lambda \in O(6,6,\mathbb{R})$,
then act on the space of states (\ref{rroperator})
via operators ${\bf \Lambda}$, whose action on
$\left( \psi^{\dag}, \psi \right) =
\left( \psi^{i \dag}, \psi_j \right)_{i,j=1, \dots , 6}$
is given by
\be
\Lambda = \left(
 \begin{array}{cc}
 A & B \\ C & D
 \end{array}
\right) :
\left( {\bf \Lambda} \psi^{\dag} {\bf \Lambda}^{-1},
 {\bf \Lambda} \psi {\bf \Lambda}^{-1} \right) =
\left( \psi^{\dag}, \psi \right)
\left(
 \begin{array}{cc}
 A & B \\ C & D
 \end{array}
\right)\ .
\ee
At the same time, the internal components of the metric (in our case of vanishing
internal NSNS $B$-field) transform according to
\be \label{etrans}
G \mapsto (AG + B) (CG + D)^{-1}\ .
\ee
We are interested in just one particular element that corresponds
to inverting the radii of all $d=6$ circles. This is in fact not a
completely well specified operation, since the signs that can
appear in the mapping of the RR forms depend on the order in which
the circles are dualized. In view of (\ref{etrans}) we choose the
element $\Lambda$ with $A=D=0$ and $B=C={\bf 1}_6$, where ${\bf
1}_6$ denotes the six-dimensional unity matrix. This element is
given by\footnote{In terms of the notation used in
\cite{hepth9907132} this corresponds to  ${\bf \Lambda} = {\bf
C^-}$.}
\be
{\bf \Lambda} = {\bf \Lambda}^-_1 \, \cdots \, {\bf \Lambda}^-_6\ , \quad
{\bf \Lambda}^-_i = \psi^{i \dag} - \psi_i\ ,
\ee
which satisfies ${\bf \Lambda}^2 = -{\bf 1}$ and
%
\be
{\bf \Lambda} \psi_i {\bf \Lambda}^{-1} = \psi^{i \dag}\ , \quad
{\bf \Lambda} \psi^{i \dag} {\bf \Lambda}^{-1} = \psi_i \ .
\ee
The action of ${\bf \Lambda}$
on a given RR form (\ref{rroperator}) can then be obtained from
\be
{\bf \Lambda} \psi^{i_1 \dag}\, \cdots\, \psi^{i_{n} \dag} \vert 0 \rangle
= \psi_{i_1}\, \ldots\, \psi_{i_{n}} {\bf \Lambda}  \vert 0 \rangle
= \frac{(-1)^{n(n-1)/2}}{(6-n)!} \hat \epsilon_{i_1 \ldots i_{6}}
\psi^{i_{n+1} \dag}\, \ldots\, \psi^{i_{6} \dag} \vert 0 \rangle\ ,
\ee
where $\hat \epsilon$ has been introduced below (\ref{tdrr2}).


\section{Kinetic and CS terms for the scalars $(C_4)_{ijkl}$}
\label{kinc4}

The kinetic terms for the scalars $(C_4)_{ijkl}$ given in (\ref{kinrr}) (however without 3-form flux)
and the Chern-Simons terms of (\ref{cs}) are also
derivable directly by a reduction of the truncated type IIB action.
In this section we calculate them in a similar fashion
as is done e.g.\ in \cite{hepth0202168,MO}. We start with
the pseudoaction of type IIB string theory without imposing the
self-duality of the 5-form field strength, reduce it to four dimensions and
afterwards impose the self-duality by adding a suitable Lagrange multiplier.
The relevant part of the ten-dimensional action is
\beqn \label{F5terms}
{\cal S} = \frac{1}{8 \kappa_{10}^2} \int
F_5 \wedge * F_5 - \frac{1}{4
\kappa_{10}^2}
\int C_4 \wedge dB_2 \wedge dC_2 + \cdots\ .
\eeqn
The field strength $F_5$ is defined as in (\ref{rrfs}),
i.e.\ after the truncation to the T-dual type I$'$ theory
\be \label{f5expand}
F_5 = (dC_4)^{\{1,4\}}
+ (dC_4)^{\{3,2\}} - \frac12 B_2^{\{1,1\}} \wedge (dC_2)^{\{2,1\}} +
\frac12 C_2^{\{1,1\}} \wedge (dB_2)^{\{2,1\}}\ .
\ee
This corresponds to expanding $C_4$ as
\be \label{c4expand}
C_4 = C_4^{\{2,2\}} + C_4^{\{0,4\}}\ .
\ee
The self-duality condition for $F_5$ gives
\be \label{selfdual}
(dC_4)^{\{1,4\}}
= * \Big( (dC_4)^{\{3,2\}} - \frac12 B_2^{\{1,1\}} \wedge (dC_2)^{\{2,1\}} +
\frac12 C_2^{\{1,1\}} \wedge (dB_2)^{\{2,1\}} \Big)\ .
\ee
Plugging (\ref{f5expand}) and (\ref{c4expand}) into (\ref{F5terms}) we obtain
\beqn \label{sns}
{\cal S} & = & \frac{1}{8 \kappa_{10}^2} \int \Big(
\Big[ (dC_4)^{\{3,2\}} - \frac12 B_2^{\{1,1\}} \wedge (dC_2)^{\{2,1\}} +
\frac12 C_2^{\{1,1\}} \wedge (dB_2)^{\{2,1\}} \Big]  \non & &
\hspace{2.2cm}  \wedge * \Big[ (dC_4)^{\{3,2\}} - \frac12
B_2^{\{1,1\}} \wedge (dC_2)^{\{2,1\}} + \frac12 C_2^{\{1,1\}} \wedge
(dB_2)^{\{2,1\}} \Big] \non & & \hspace{1.6cm}
+ (dC_4)^{\{1,4\}}
\wedge * (dC_4)^{\{1,4\}}
\Big) \non
& &
\mbox{} - \frac{1}{4 \kappa_{10}^2} \int
C_4^{\{0,4\}} \wedge (dB_2)^{\{2,1\}} \wedge (dC_2)^{\{2,1\}}\ .
\eeqn
In order to implement the self-duality condition (\ref{selfdual}) we add the
Lagrange multiplier $\check C_4^{\{0,4\}}$
%
%
\be \label{lagmulti}
\delta {\cal S} = - \frac{1}{4\kappa^2_{10}}
 \int \Big( F_5^{\{3,2\}} + \frac12 B_2^{\{1,1\}} \wedge (dC_2)^{\{2,1\}}
- \frac12 C_2^{\{1,1\}} \wedge (dB_2)^{\{2,1\}} \Big) \wedge (d\check C_4)^{\{1,4\}}\ .
\ee
The equation of motion for $\check C_4^{\{0,4\}}$ implies
\be
F_5^{\{3,2\}} = (dC_4)^{\{3,2\}} - \frac12 B_2^{\{1,1\}} \wedge (dC_2)^{\{2,1\}} +
\frac12 C_2^{\{1,1\}} \wedge (dB_2)^{\{2,1\}}\ .
\ee

On the other hand, the equation of motion for $F_5^{\{3,2\}}$ in combination
with the self-duality condition (\ref{selfdual}) leads to the identification of
$(dC_4)^{\{1,4\}}$ and $(d\check C_4)^{\{1,4\}}$
\be
(d\check C_4)^{\{1,4\}} = (dC_4)^{\{1,4\}}\ .
\ee
Using this in ${\cal S} + \delta {\cal S}$ gives (after a partial
integration)
\be \label{ausint}
{\cal S} = \frac{1}{4\kappa^2_{10}}  \int (dC_4)^{\{1,4\}}
\wedge * (dC_4)^{\{1,4\}}
 + \frac{1}{2\kappa^2_{10}}
 \int  (dC_2)^{\{2,1\}}  \wedge (dB_2)^{\{2,1\}} \wedge C_4^{\{0,4\}}\ ,
\ee
which obviously coincides with (\ref{cs}) and the kinetic term for $C_4^{\{0,4\}}$
in (\ref{kinrr}) in the absence of fluxes. \\

Let us also mention here that it is not straightforward to modify
this procedure to include also the 3-form fluxes. If one just
naively plugs the ansatz with 3-form fluxes into the action
(\ref{F5terms}), one has to replace
\be \label{replace}
(dC_4)^{\{1,4\}} ~\rightarrow~ (dC_4)^{\{1,4\}} -
\frac12 B_2^{\{1,1\}} \wedge (dC_2)^{\{0,3\}} + \frac12 C_2^{\{1,1\}}
\wedge (dB_2)^{\{0,3\}}
\ee
in (\ref{f5expand}), (\ref{selfdual}) and (\ref{sns}). In addition
one gets a further Chern-Simons term
\be \label{addcs}
- \frac{1}{4 \kappa_{10}^2} \int C_4^{\{2,2\}} \wedge \Big(
(dB_2)^{\{2,1\}} \wedge (dC_2)^{\{0,3\}} - (dC_2)^{\{2,1\}} \wedge
(dB_2)^{\{0,3\}} \Big)\ .
\ee
One immediately runs into trouble now when one
considers the equation of motion for $C^{\{2,2\}}$. It is given by
\be \label{c22eom}
d^{\{1,0\}} * F_5^{\{3,2\}} = -(dB_2)^{\{2,1\}}
\wedge (dC_2)^{\{0,3\}} + (dC_2)^{\{2,1\}} \wedge (dB_2)^{\{0,3\}}\ ,
\ee
which, however, is not consistent with the self-duality condition
(\ref{selfdual}) after performing the substitution
(\ref{replace}). This would require a factor 1/2 on the right side
of (\ref{c22eom}), as
\be
d^{\{1,0\}} F_5^{\{1,4\}} = - \frac12 (dB_2)^{\{2,1\}} \wedge (dC_2)^{\{0,3\}} +
 \frac12 (dC_2)^{\{2,1\}} \wedge (dB_2)^{\{0,3\}}\ .
\ee
This formula is consistent with the $\{2,4\}$-component of the Bianchi identity
\be
(dF_5)^{\{2,4\}} = -(dB_2)^{\{2,1\}} \wedge (dC_2)^{\{0,3\}} +
(dC_2)^{\{2,1\}} \wedge (dB_2)^{\{0,3\}} \ ,
\ee
as the latter actually gets two contributions. Only
one is from $d^{\{1,0\}} F_5^{\{1,4\}}$. The other one comes from
$d^{\{0,1\}} F_5^{\{2,3\}}$ and
is identical to the first one.

\section{The self-dual type I action}
\label{sdactionsec}

Due to the fact that the conventional formulation of the CS action of D-branes involves
RR potentials of all (even) degrees (in type IIB) \cite{hepth9910053}, it might seem more
appropriate to start with a formulation of type I supergravity that is democratic
in the appearance of the RR fields $C_2$ and $C_6$. Its closed string part could be
obtained by quotienting world-sheet parity out of the democratic version of type IIB,
whose RR sector was given in (\ref{iib2}), and which is also much better adapted to
make the invariance of the type IIB action under T-duality explicit. We did not follow this approach
because the introduction of the superfluous degrees of freedom is only needed
in a discussion of the correct treatment of the CS-terms appearing in the D3-brane
action, cf.\ section \ref{secaddmod}. For the derivation of the T-dual action
performed in chapter \ref{sec3} it is actually not necessary to work with this democratic
formulation. Nevertheless, we give it here for completeness and because of its relevance for
the discussion in section \ref{secaddmod}.
\\

The self-dual type I action which contains $C_2$ and $C_6$
democratically is given by\footnote{Actually, as already mentioned in the
main text, in the presence of the $\omega_7$-terms it is not consistent to just keep
the quadratic term $|{\cal F}|^2$ in an expansion of the DBI action. It is understood that
further terms would have to be added in a rigorous treatment.}
\beqn \label{sdaction}
&& {\cal S}_{\rm self-dual} =
\frac{1}{2 \kappa_{10}^2} \int d^{10} x\, \sqrt{-g} \Big(
  e^{-2 \Phi} \left(R
  + 4 \partial_\mu \Phi \partial^\mu \Phi \right)
  - \frac14 |\tilde F_3|^2
  - \frac14 |\tilde F_7|^2 \Big) \non
&&\hspace{2cm}
  - \frac{1}{2 g_{10}^2} \int d^{10} x\, \sqrt{-g}
e^{-\Phi} {\rm tr}\, |{\cal F}|^2  \\
&&\hspace{2cm}
-  \frac{1}{4 g_{10}^2} \int dC_2 \wedge \omega_7
- \frac{1}{4 g_{10}^2} \int dC_6 \wedge \omega_3
+  \frac{\kappa_{10}^2}{4 g_{10}^4}
  \int \omega_7 \wedge \omega_3\ .
\nonumber
\eeqn
We have included terms
$dC_p \wedge \omega_{9-p}$ stemming
from the Chern-Simons action $C_p \wedge {\rm ch}({\cal F})$
of the 9-branes.
In analogy to $\tilde F_3$ we then define
\be \label{fptilde}
\tilde F_7 = dC_6 -
   \frac{\kappa_{10}^2}{g_{10}^2} \omega_7\ .
\ee
The normalization of the D9-brane CS terms is exactly one half of the ordinary.
In addition to this action the duality condition
\be \label{duality}
* \tilde F_3
= - \tilde F_7
\ee
has to be imposed after deriving the equations of motion.
This procedure is justified primarily by checking that the
equations of motion and Bianchi identities that result
from the self-dual action
become identical to the standard equations derived from
(\ref{iaction}) plus the
D9-brane interaction $dC_2 \wedge \omega_7$, after imposing the
constraint (\ref{duality}). This requirement lead us to introduce the slightly exotic
interaction $\omega_7 \wedge \omega_3$.\footnote{Such terms are, however, familiar from
anomaly cancellation, see e.g.\ formula (A.14) in \cite{DM}.} \\

Let us also check that the action (\ref{sdaction})
really reproduces the old action (\ref{iaction}) plus
D-brane Chern-Simons term after the duality constraint (\ref{duality})
has been eliminated.
Thus we add the Lagrange multiplier term
\be \label{Lagrm}
\delta {\cal S} = -\frac{1}{4 \kappa_{10}^2} \int d \check C_2 \wedge dC_6
= -\frac{1}{4 \kappa_{10}^2} \int d \check C_2 \wedge
(\tilde F_7 +  \frac{\kappa_{10}^2}{g_{10}^2} \omega_7 )\ .
\ee
The equation of
motion for $\tilde F_7$ is just
\be \label{dual}
* \tilde F_7 = - \check F_3 =
- d \check C_2 + \frac{\kappa_{10}^2}{g_{10}^2} \omega_3 \ ,
\ee
i.e.\ $\check F_3$ has the same dependence on $\check C_2$ as $\tilde F_3$ has on $C_2$.
Comparing with (\ref{duality}) and using $** = 1$ for a
form of odd degree in an even-dimensional space-time of
Lorentzian signature, we see that
\be \label{hftf}
\check F_3 = \tilde F_3\ .
\ee
Using this and (\ref{dual}) to eliminate $\tilde F_7$ leads to the
original action
\beqn \label{sdual1}
{\cal S}_{\rm I}'
& = & \frac{1}{2 \kappa_{10}^2} \int d^{10} x\, \sqrt{-g} \left( e^{-2 \Phi}
 \left( R + 4 \partial_\mu \Phi \partial^\mu \Phi \right)
- \frac12 |\tilde F_3|^2 \right) \non
&&
- \frac{1}{2 g_{10}^2} \int d^{10} x\, \sqrt{-g}
   e^{-\Phi} {\rm tr}\, |{\cal F}|^2
- \frac{1}{2 g_{10}^2} \int dC_2 \wedge \omega_7
\eeqn
involving only the RR 2-form $C_2$ including
the properly normalized Chern-Simons action of a D9-brane. \\

On the other hand, one may want to
integrate out $C_2$ in place of $C_6$. Instead of (\ref{Lagrm}) we add
\be \label{lagm}
\delta {\cal S} = -\frac{1}{4 \kappa_{10}^2} \int d C_2 \wedge d \check C_6
= -\frac{1}{4 \kappa_{10}^2} \int (\tilde F_3
+ \frac{\kappa_{10}^2}{g_{10}^2} \omega_3) \wedge d \check C_6 \ .
\ee
The equation of motion of $\tilde F_3$ is given by
\be \label{dualdual}
* \tilde F_3 = d \check C_6 +
  \frac{\kappa_{10}^2}{g_{10}^2} \omega_7\ .
\ee
Comparing this with (\ref{duality}) now leads to the relation
\be \label{chc}
d \check C_6 = - d C_6\ .
\ee
With the help of (\ref{chc}) and (\ref{dualdual}) we can eliminate
$\tilde F_3$ and get
\beqn \label{sdual2}
{\cal S}_{{\rm I}}''
& = &\frac{1}{2 \kappa_{10}^2} \int d^{10} x\, \sqrt{-g}
\left( e^{-2 \Phi} \left( R + 4 \partial_\mu \Phi \partial^\mu \Phi \right)
- \frac12 |\tilde F_7|^2  \right)
\non
&&
- \frac{1}{2 g_{10}^2} \int d^{10} x\, \sqrt{-g}
   e^{-\Phi} {\rm tr}\, |{\cal F}|^2
- \frac{1}{2 g_{10}^2} \int \tilde F_7 \wedge \omega_3
\eeqn
In particular there is now an extra Chern-Simons term of the form
\be \label{csterm}
\frac{1}{2 g_{10}^2} \int \omega_3 \wedge \omega_7 \ .
\ee
%


\end{appendix}

\end{document}